# Disorder-induced enhancement of lithium-ion transport in solid-state electrolytes

Zhimin Chen[1], Tao Du[1,*], N. M. Anoop Krishnan[2], Yuanzheng Yue[1], Morten M. Smedskjaer[1,*]

[1] *Department of Chemistry and Bioscience, Aalborg University, Aalborg East 9220, Denmark*
[2] *Department of Civil Engineering, Indian Institute of Technology Delhi, New Delhi 110016, India*
* *Corresponding authors. E-mail:* taod@bio.aau.dk *(T.D.),* mos@bio.aau.dk *(M.M.S.)*

**ABSTRACT**
Enhancing the ion conduction in solid electrolytes is critically important for the development of high-performance all-solid-state lithium-ion batteries (LIBs). Lithium thiophosphates are among the most promising solid electrolytes, as they exhibit superionic conductivity at room temperature. However, the lack of comprehensive understanding regarding their ion conduction mechanism, especially the effect of structural disorder on ionic conductivity, is a long-standing problem that limits further innovations of all-solid-state LIBs. Here, we address this challenge by establishing and employing a deep learning potential to simulate $Li_3PS_4$ electrolyte systems with varying levels of disorder. The results show that disorder-driven diffusion dynamics significantly enhances the room-temperature conductivity. We further establish bridges between dynamical characteristics, local structural features, and atomic rearrangements by applying a machine learning-based structure fingerprint termed "softness". This metric allows the classification of the disorder-induced "soft" hopping lithium ions. Our findings offer insights into ion conduction mechanisms in complex disordered structures, thereby contributing to the development of superior solid-state electrolytes for LIBs.

**Introduction**

Lithium-ion batteries (LIBs) have revolutionized portable electronics and play an increasingly important role in electric vehicles and grid energy storage due to their high energy density and long cycle life[1–3]. However, as demands for higher energy density, enhanced safety, and faster charging continue to grow, traditional LIBs are rapidly approaching their performance limits[4–6], underscoring the urgent need to explore new frontiers in battery technology. One of the most promising avenues is the development of solid-state batteries, in which the liquid electrolyte is replaced with a solid electrolyte, enabling safer, more efficient, and extended lifespan in energy storage solutions. Unlike their liquid counterparts, solid electrolytes are non-flammable and intrinsically safer[7], mitigating the risk of thermal runaway events[8,9], leakage[10], and chemical instability[11]. Glassy solid electrolytes are interesting candidates for solid-state batteries, considering their various advantages over the crystalline counterparts such as isotropic ion conduction[12], lack of grain boundaries[13], and ease of industrial processing[14]. Among these solid-state electrolytes, lithium thiophosphate based glasses and glass-ceramics are especially promising due to their high ionic conductivity[15–17], minimal ion transfer resistance at the electrode interface, and cost-effective processing[18].

The mechanism of ion conduction within solid-state electrolytes is fundamentally different from that of liquid electrolytes. In liquid electrolytes, ions move through the liquid medium and electrons are separated[19], while in solid-state electrolytes, ions diffuse along favorable migration pathways in crystalse[7] or navigate a disordered structural landscape[20]. Understanding the intricacies of ion conduction in solid-state electrolytes is critical for optimizing battery performance. The mechanisms of ion conduction in crystalline solid electrolytes include vacancy-assisted migration, interstitial diffusion, and even tunneling[7]. However, the mechanisms for the ion conduction in disordered or glassy solid electrolytes is yet to be established[7,19,21], primarily because of the inherently irregular nature of the energy landscape in these materials[20]. Furthermore, the lack of a well-defined crystal lattice makes it challenging to predict and control ion pathways accurately. Additionally, the presence of defects and disorder-induced structural heterogeneity further complicates the understanding of ion conduction in these materials.

The use of large-scale molecular dynamics (MD) simulations allows for capturing the long-term (up to hundreds of nanoseconds) dynamic changes of hopping ions. However, to the best of our knowledge, there is currently no potential function available for simulating processes like the glass transition and phosphorus-sulfur (P-S) bond breaking in the important family of lithium phosphorus sulfide (LiPS) systems. It is worth noting that the recent classical potential proposed by Ariga et al.[22] has certain limitations, particularly on bond breaking and reactions. Similarly, the classical potential proposed by Kim et al.[23] not only prohibits bond breaking but is also limited to γ-$Li_3PS_4$. To address this, we here train a machine learning-based interatomic potential (MLIP)[24] based on *ab initio* molecular dynamics (AIMD) training data. As shown herein, this new potential allows for the simulation of both crystalline and glassy forms of LiPS with an accuracy comparable with AIMD simulations but at a much lower computational cost. With the objective to unravel the structural origins of disordered-induced acceleration of lithium-ion migration, we use the MLIP to construct glassy $Li_3PS_4$ electrolytes, as well as ordered β-$Li_3PS_4$ (*Pnma*) and partially crystalline $Li_3PS_4$ glass-ceramics (i.e., three systems with varying degree of disorder), and quantify structural descriptors for order and disorder. Further, we compare the homogeneous dynamics (mean-squared displacement, van Hove correlation functions) and heterogeneous dynamics (non-Gaussian statistics) across the systems with varying degrees of disorder. We then focus on the partially crystalline Li$_3$PS$_4$ glass-ceramic electrolyte, quantifying the dynamic distinctions between its internal order and disorder, including interfaces. Finally, we employ the classification-based structure fingerprint termed softness[25–27] to identify disorder-induced soft hopping ions. Taken as a whole, the present investigation of hopping ions' conduction mechanisms in solid-state electrolytes reveal the origin of disorder-driven fast transport of lithium ions. Notably, the combined approach involving molecular dynamics and machine learning is broadly applicable to studying other types of solid-state ion transport systems.

**Structure descriptors for order and disorder.** To strike a balance between accuracy and computational efficiency, we develop a MLIP for the LiPS system. The MLIP is trained using trajectories obtained from DFT-level AIMD simulations as the dataset (see *Methods* for details). The accuracy of the MLIP in reproducing structural information obtained from AIMD is shown in **Fig. 1a**. The DeePMD based potential demonstrates an almost DFT-level accuracy in reproducing short-range structural features, closely matching the first and second coordination shell in the $g(r)$. Moreover, compared to MD simulations using a classical potential[22], which also cannot capture bond breaking and reaction events, the present MLIP-based simulations consistently exhibit a better agreement with the AIMD results. Additionally, we provide a comparative analysis with experimental scattering data. **Fig. S1** shows the comparison between MD simulation and neutron scattering data regarding the structure factor $S(Q)$ for glassy Li$_3$PS$_4$. The $R_\chi$ factors in **Fig. S1**, as introduced by Wright[28], are calculated, confirming the agreement between the simulated and experimental scattering results, demonstrating a good agreement for the present MLIP.

**Fig. 1b** presents the contrasting spatial distributions of $PS_4^{3-}$ units within both glassy Li$_3$PS$_4$ (top) and crystalline β-Li$_3$PS$_4$ (bottom). The disordered arrangement of $PS_4^{3-}$ units in the glass matrix is charge-balanced by lithium ions. We find no regular coordination sites and symmetric long-range migration pathways within the glassy Li$_3$PS$_4$ electrolyte[7], implying isotropic ionic conduction[19]. This differs notably from its crystalline counterparts. In the β-Li3PS4 structure, the $PS_4^{3-}$ motifs adopt an ordered zig-zag arrangement, accommodating lithium ions within both octahedral and tetrahedral interstitials[29]. These interstitials allow for discernible migration paths along designated transport channels for the lithium ions[30].

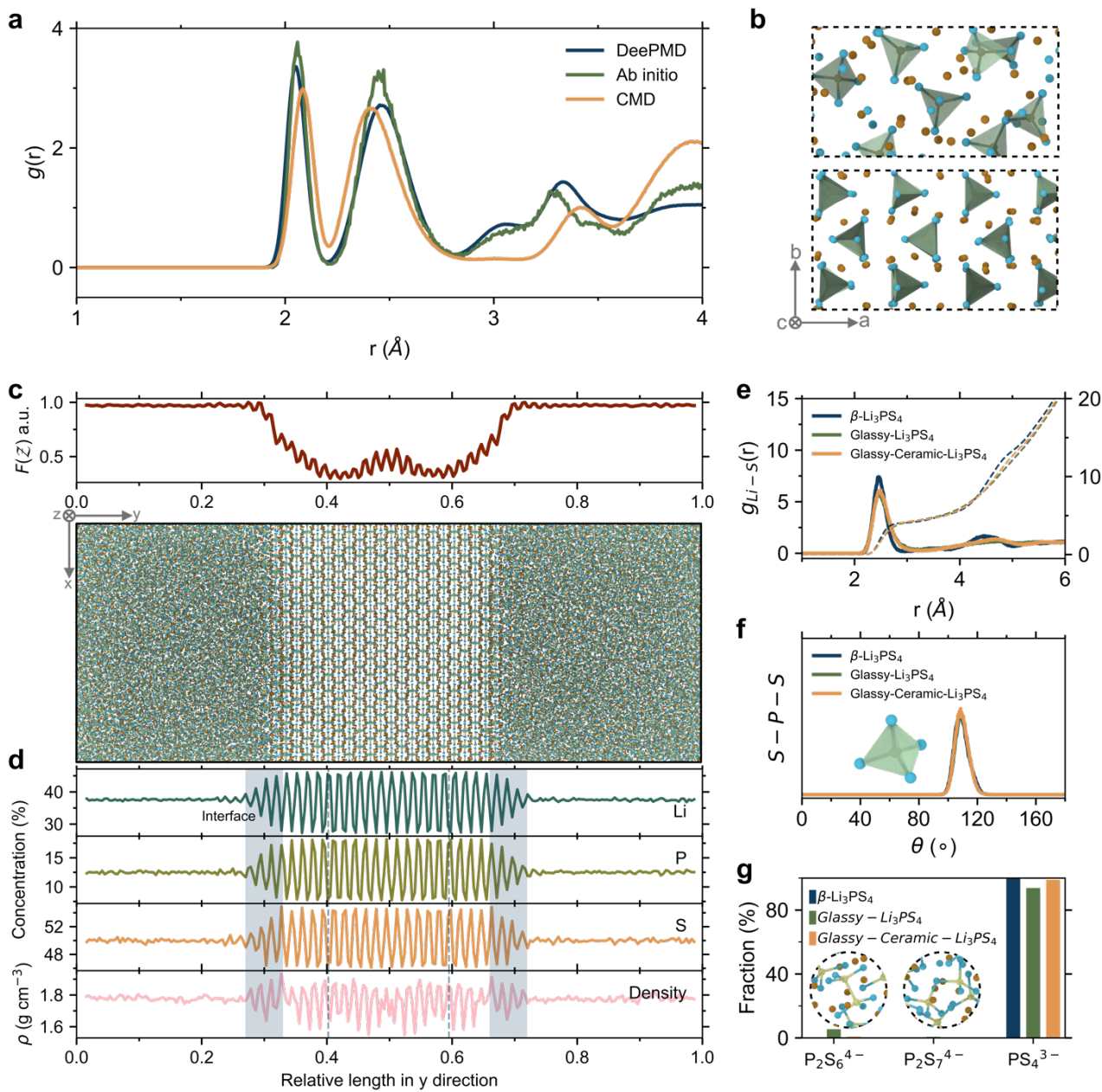

**Fig. 1 Structural fingerprints in the transition from ordered to disordered electrolytes. a** Pair distribution function $g(r)$ of β-$Li_3PS_4$ electrolytes simulated by molecular dynamics simulations using machine learning interatomic potential (MLIP), DFT-based ab initio molecular dynamics (AIMD) simulations, and classical molecular dynamics (CMD) simulations[22]. **b** Atomic snapshots of glassy $Li_3PS_4$ (top) and β-$Li_3PS_4$ (bottom) electrolyte configurations. The snapshots are captured from localized regions within the final relaxed configurations in the MD simulation with the MLIP. **c** Atomic snapshot of the glass-ceramic $Li_3PS_4$, depicting the amorphization distribution (**eq. 6**) of the constructed glass-ceramic along the *y*-axis at the top. **d** Element concentration and density profiles for the glass-ceramic $Li_3PS_4$ from panel **c**. The gray rectangle and dashed line highlight the internal interface between the ordered and disordered phases and the exemplified crystalline plane. **e** Radial distribution function (RDF) of Li-S pairs and integrated RDFs for glassy-, β-, and glass-ceramic $Li_3PS_4$. **f** Angular distribution function of S-P-S. **g** Fractions of thiophosphate anions in glassy-, β-, and glass-ceramic $Li_3PS_4$ system.

**Fig. 1c** shows snapshots of the glass-ceramic sample Li$_3$PS$_4$ (see *Methods* for details on the construction of the glass-ceramic model using MD simulations), and a visual representation of the amorphization $F(\mathcal{Z})$ distribution along the y-axis of simulation cell. $F(\mathcal{Z})$ is calculated from the Gaussian density of the atomic arrangement through the Fourier transform method, as explained in the *Methods* section and Ref.[31]. The glass phase has a $F(\mathcal{Z})$ value of approximately 1, with a transition of the degree of disorder occurring at the interface between the glass and crystal phases within the $F(\mathcal{Z})$ profile, that is, the transition zone where the curve exhibits a double-minima. Corresponding profiles of element concentration and density are depicted in **Fig. 1d** for the glass-ceramic Li$_3$PS$_4$ electrolyte. We compute averages along the *y*-axis using specific bin values derived from simulated boxes. The resulting profiles reveal the homogeneity of the glassy phase, whereas elements in the crystalline phase display a periodic distribution. Notably, the gray rectangle and dashed line highlight the internal interface where the ordered and disordered phases coexist and an exemplified crystalline plane, respectively.

The glassy, crystalline, and glass-ceramic Li$_3$PS$_4$ are three solid-state electrolytes with distinctly different structures, i.e., varying degree of disorder. Their short-range order (SRO) and medium-range order (MRO) structures are characterized by the radial distribution function (RDF) $g(r)$ and structure factor $S(Q)$, respectively. As shown in **Fig. S2**, a pronounced first sharp diffraction peak (FSDP) of neutron $S_N(Q)$ is observed in β-Li$_3$PS$_4$, indicating a higher degree of structural order. Additionally, in reciprocal space, the β-Li$_3$PS$_4$ exhibits a peak at lower $Q$-values compared to that in the glassy Li$_3$PS$_4$, signifying that its structural order extends over a longer length scale than that of glassy Li$_3$PS$_4$. **Fig. S3** depicts the SRO structure of Li$_3$PS$_4$, showcasing a comparable distribution of P-S distances in the first peaks of both the disordered and ordered Li$_3$PS$_4$. Furthermore, the first peak is located at approximately 2.07 Å, corresponding to the length of P-S bonds[32,33]. We also discern a discernible periodicity in the distribution of Li-Li distances within the β-Li$_3$PS$_4$ sample.

The RDF of Li-S pairs is shown in **Fig. 1e**, offering insights into the interactions of ions within Li$_3$PS$_4$ systems. Additionally, the integrated RDFs for glassy Li$_3$PS$_4$, crystalline β-Li$_3$PS$_4$, and glass-ceramic Li$_3$PS$_4$ are compared. Despite the structural transition from order to disorder, the distance distributions of Li and S remain largely consistent, with a coordination number of approximately 4 for Li. In β-Li$_3$PS$_4$, Li is situated within both tetrahedral (LiS$_4$) and octahedral interstitial sites (LiS$_6$) among the S atoms. Conversely, in the disordered structure, the Li atoms in the four-coordinated configuration occupy a larger free volume (see **Fig. S4**). The angular distribution function of S-P-S is shown in **Fig. 1f**, revealing the tetrahedral PS$_4^{3-}$ motifs formed between sulfur and phosphorous atoms. This provides insights into the local bonding states within the system. In glassy and glass-ceramic Li$_3$PS$_4$ electrolytes, the emergence of the P-P peak around 2.3 Å (**Fig. S3**) is attributed to the formation of thermodynamically more stable P$_2$S$_6^{4-}$ units during the melt-quenching process of the disordered structure[34,35]. We quantify the occurrence of these units in various systems, and as depicted in **Fig. 1g**, both P$_2$S$_6^{4-}$ and P$_2$S$_7^{4-}$ units are detected in both glassy and glass-ceramic Li$_3$PS$_4$, with their presence exclusively observed within the disordered glass phase in the latter.

**Dynamics of hopping ions.** The impact of disorder on lithium-ion mobility and transport properties is investigated by analyzing the dynamic diffusion behavior of hopping ions within the Li$_3$PS$_4$ systems. We first evaluate the time-averaged mean-squared displacement (MSD, $\overline{\langle r^2(t) \rangle}$) of lithium ions (**Fig.**

**2a**). The MSD quantifies the average distance the ions travel over time, and at high temperatures, the β-$Li_3PS_4$ electrolytes demonstrate the highest MSD at 900 K (**Fig. 2a**) and 1000 K (**Figs. S5a-c**). Conversely, the glass-ceramic $Li_3PS_4$, comprising both glassy and crystalline phases, exhibits a marginally reduced MSD compared to the glassy $Li_3PS_4$. All $Li_3PS_4$ electrolytes feature diffusion dynamics at 1000 K, eventually reaching Fick's limit ($t^1$), as indicated by the exponent of MSD in **Figs. S5d-e** (β-$Li_3PS_4$ reaches $t^1$ at 900 K as shown in **Fig. S5d**). As the temperature decreases, the cumulative displacement difference of the hopping lithium ions between β-$Li_3PS_4$ and both the glassy $Li_3PS_4$ and glass-ceramic $Li_3PS_4$ diminishes gradually. Specifically, the $Li^+$ MSD for β-$Li_3PS_4$ decreases with decreasing temperature, ultimately becoming smaller than that of both glassy and glass-ceramic $Li_3PS_4$ at 700 K (**Figs. S5d-e**). Interestingly, the disordered structure, characterized by isotropic transport pathways, exhibits the smaller decrease in MSD with decreasing temperature compared to the more ordered structure. The latter possesses a predominant transport pathway connecting octahedral and tetrahedral interstitials[30].

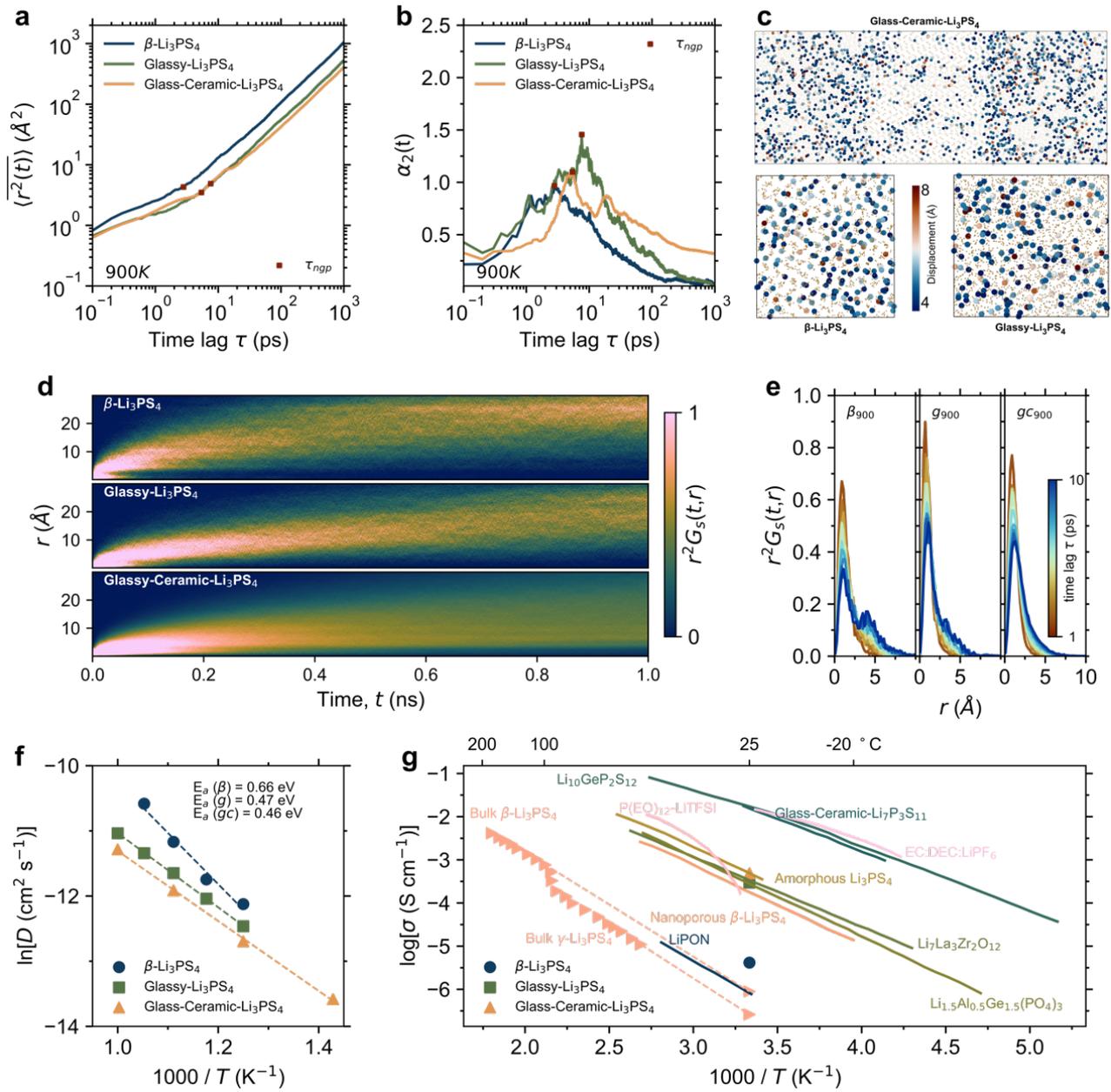

**Fig. 2 Disorder-driven fast lithium diffusion. a** Time-averaged mean-squared displacement (MSD, $\overline{\langle r^2(t) \rangle}$), and **b** non-Gaussian parameter (NGP, $\alpha_2(t)$) of lithium ions for β-, glassy-, and glass-ceramic Li$_3$PS$_4$ systems, as functions of the time lag, $\tau$. The red squares represent the NGP peak times, $\tau_{ngp}$. **c** Short-term time-averaged displacement of lithium ions for 10 ps. The displacement magnitudes exceeding 4 Å are highlighted with larger markers for better representation. **d** Self-part van Hove correlation function for β-, glassy-, and glass-ceramic Li$_3$PS$_4$ systems at 900 K. **e** Self-part van Hove correlation function for β-, glassy-, and glass-ceramic Li$_3$PS$_4$ systems for fixed time between 1 and 10 ps. In panel **e**, the abbreviations $β_{900}$, $g_{900}$, and $gc_{900}$ refer to β-, glassy-, and glass-ceramic Li$_3$PS$_4$ at 900K. **f** Arrhenius fit of the temperature dependence of the lithium diffusion coefficient. **g** Comparison of temperature-dependence of ionic conductivity in Li$_3$PS$_4$ systems with varying degree of disorder, as well as results for other lithium solid electrolytes, organic liquid electrolytes, and polymer electrolytes[17,36–44].

Particles may exhibit anomalous diffusion, where their displacement distribution deviates from a Gaussian distribution. This illustrates the situation where particles within one domain exhibit notably faster movement compared to those in neighboring domains, typically separated by a few nanometers. This phenomenon is commonly referred to as dynamic heterogeneity[45]. For a more comprehensive understanding of the temporal dynamical events involving lithium-ions, we therefore employ the non-Gaussian parameter (NGP) $\alpha_2(t)$ descriptor to compare the anomalous diffusion within the different $Li_3PS_4$ systems. The NGP, or the fourth cumulant of displacement, has proven to be a measure of diffusion coefficient fluctuations and dynamic heterogeneity[46]. It characterizes the degree of deviation from the Gaussian behavior in particle diffusion. As shown in **Fig. 2b**, the motion of lithium ions exhibits non-Gaussian properties, indicating that it is not entirely random. Red squares denote the NGP peak times, symbolized as $\tau_{ngp}$, signifying the occurrence of non-Gaussian dynamics. An increase in the degree of disorder leads to a delay in the emergence of the NGP peaks and an increase in their peak heights. As illustrated in **Fig. S6**, both the NGP peak height and time increase as the temperature decreases. We focus on the short-term (10 ps) variations in lithium-ion displacement by enhancing the visualization of significant motions, as shown in **Fig. 2c**, where larger atoms are used to emphasize displacement magnitudes exceeding 4 Å. In the disordered glassy phase, there is an excess of lithium ions rapidly deviating from their initial vibrational positions, covering longer displacements compared to those in the crystalline phase.

To further explore the correlation between particle mobility for short-to-long term behavior, **Figs. 2d-e** showcase the self-part van Hove correlation function ($G_s$) of Li-Li for the different $Li_3PS_4$ system. $G_s(r,t)$ characterizes the Li-Li pair distance $r$ at time $t$, and the quantity $r^2G_s(r,t)$ describes the probability distribution of particle displacements[47]. As seen from **Fig. 2d**, β-$Li_3PS_4$ has higher probability distributions for long-distance displacements (>10 Å) with increasing time. This can be attributed to the extended displacements of hopping ions along specific transport pathways within the crystalline structure[30]. In contrast, glassy $Li_3PS_4$ performs less favorably due to the convoluted migration paths of ions within its disordered structure. Surprisingly, the glass-ceramic $Li_3PS_4$ does not display a displacement distribution that lies between glass and crystal $Li_3PS_4$. This can be attributed to the orientation of the crystalline phases, as hopping ions exhibit a preferential diffusion path along the *c*-direction on the *ac*-plane in β-$Li_3PS_4$[30] (i.e., the *z*-direction shown in **Fig. 1c**). In our analysis of the MSD within β-$Li_3PS_4$ electrolytes, we observe that the component along the *z*-direction is superior to that along the *y*-direction and significantly outperforms the *x*-direction. (**Fig. S7**). When examining the probability distribution over a fixed time interval between 1 and 10 ps (**Fig. 2e**), the glass-ceramic $Li_3PS_4$ displays a single, wide peak distribution, centered at approximately 1 Å. This distribution is predominantly influenced by the equilibrium vibrations and occupation of nearest-neighbor sites[47]. Notably, within a time interval of only 10 ps, β-$Li_3PS_4$ undergoes a transition from a single peak to a double peak, indicating that lithium ions depart from their initial equilibrium positions to initiate migration. A subtle shift of the $r^2G_s(r,t)$ distribution is also observed in the glassy $Li_3PS_4$.

**Ionic conductivity.** Macroscopic ionic conduction is the result of the collective ion migration dynamics within a system,[19] Such dynamics depends on the structure of that system. In this context, here we illustrate the impact of the transition from ordered to disordered structure on the mobility of lithium ions and, consequently, on the derived ionic conductivity. We begin with the diffusion coefficient *D*, which can be calculated as the slope of the MSD-time curve (see **Fig. 2a**). In **Fig. 2f**, the temperature dependence of the diffusion coefficient is illustrated, following an Arrhenius-type behavior, with the degree of disorder influencing the temperature dependence. This is seen from the lower diffusion activation energy ($E_a$, as determined from the slope of the plot) in glassy and glass-ceramic $Li_3PS_4$ electrolytes as compared to that in β-$Li_3PS_4$. Interestingly, glass-ceramic $Li_3PS_4$ electrolytes, which is characterized partially ordered structures, demonstrate $E_a$ and ionic conductivities (as illustrated in **Fig. 2g**) comparable to, or even slightly superior to, those of purely glassy $Li_3PS_4$. The ordering induced enhancement of ionic conductivity agrees with previous findings reported for both $Li_3PS_4$ system[16,48,49] and the $Li_2S$-$P_2S_5$ and $Li_7P_3S_{11}$ systems[15,17,50–52]. This agreement arises from the interplay of glass-phase-induced and interfacial-phase-induced (disordered structures) mechanisms in glass-ceramics, as discussed in detail below.

**Fig. 2g** compares the temperature-dependent ionic conductivity of the present $Li_3PS_4$ systems, which is calculated by means of the Nernst-Einstein equation (as described in the ***Methods*** section), with that of other types of ordered and disordered electrolytes. The simulated ionic conductivities of $Li_3PS_4$ with varying degrees of disorder align closely with previously reported experimental values[36–38]. The disordered $Li_3PS_4$ electrolytes exhibit notably high room-temperature ionic conductivity, i.e., higher than that of β-$Li_3PS_4$ electrolytes.

**Ion transport in glass-ceramic structures.** We now focus on ion hopping dynamics in the glass-ceramic $Li_3PS_4$ electrolyte, which consists of both the glassy and crystalline phases, as well as the complex interface region. **Fig. 1** illustrates the interface involving a disordered transition region of $F(\mathcal{Z})$ and density variation divisions. These three phases exhibit varying degree of disorder, which can be quantified using the amorphization function $F(\mathcal{Z})$, and in simple terms, the degree of disorder follows the sequence: glass > interface > crystalline. From the MSD data at 900 K shown in **Fig. 3a**, we find that both the glass phase and the interface outperform the crystalline phase significantly. The temperature sensitivity of the MSD in the crystal phase (**Fig. S8**) reveals that as temperature decreases, the MSD significantly drops as also observed for β-$Li_3PS_4$. The heterogeneous dynamics characterized by NGP is presented in **Fig. 3b**. An increase in disorder degree is associated with an increase in NGP peak intensity. Combining this observation with the findings presented in **Fig. 2b**, we infer that an increase in degree of disorder leads to a more pronounced deviation of ion dynamics from a Gaussian distribution. It is evident that in the glassy (**Fig. 3c**) and interfacial (**Fig. S9a**) regions of the glass-ceramic $Li_3PS_4$, there is a significant long-range displacement of hopping ions over time at 900 K. In contrast, the crystalline phase of the glass-ceramic does not exhibit the probability distribution of β-$Li_3PS_4$. Instead, it demonstrates a high probability of displacements within the range of 3-10 Å throughout the entire simulation period (highlighted as the bright regions in the bottom panel of **Fig. 3c**). Instead, it consistently exhibits a notable likelihood of displacements within the range of 3-10 Å throughout the simulation period. This suggests that lithium ions within the crystalline phase of the glass-ceramic (~8 nm in size as shown in **Fig. S9b**) cannot readily diffuse across the interface and/or migrate into the glass phase during thermal activation. However, the enhanced ionic conductivity observed in glass-ceramic $Li_3PS_4$ electrolytes, in contrast to both glassy $Li_3PS_4$ and β-$Li_3PS_4$, is closely linked to the dynamics of conducting ions driven by internal disorder within both the glassy phase and the disordered interface. Furthermore, this enhancing effect consistently occurs in glass-ceramic $Li_3PS_4$ samples (see **Fig. S10a**). Specifically, the room-temperature ionic conductivities of glass-ceramic $Li_3PS_4$ electrolytes, varying in crystal contents, exceed that of glassy $Li_3PS_4$ electrolyte. Additionally, they are more than two orders of magnitude greater than that of β-$Li_3PS_4$.

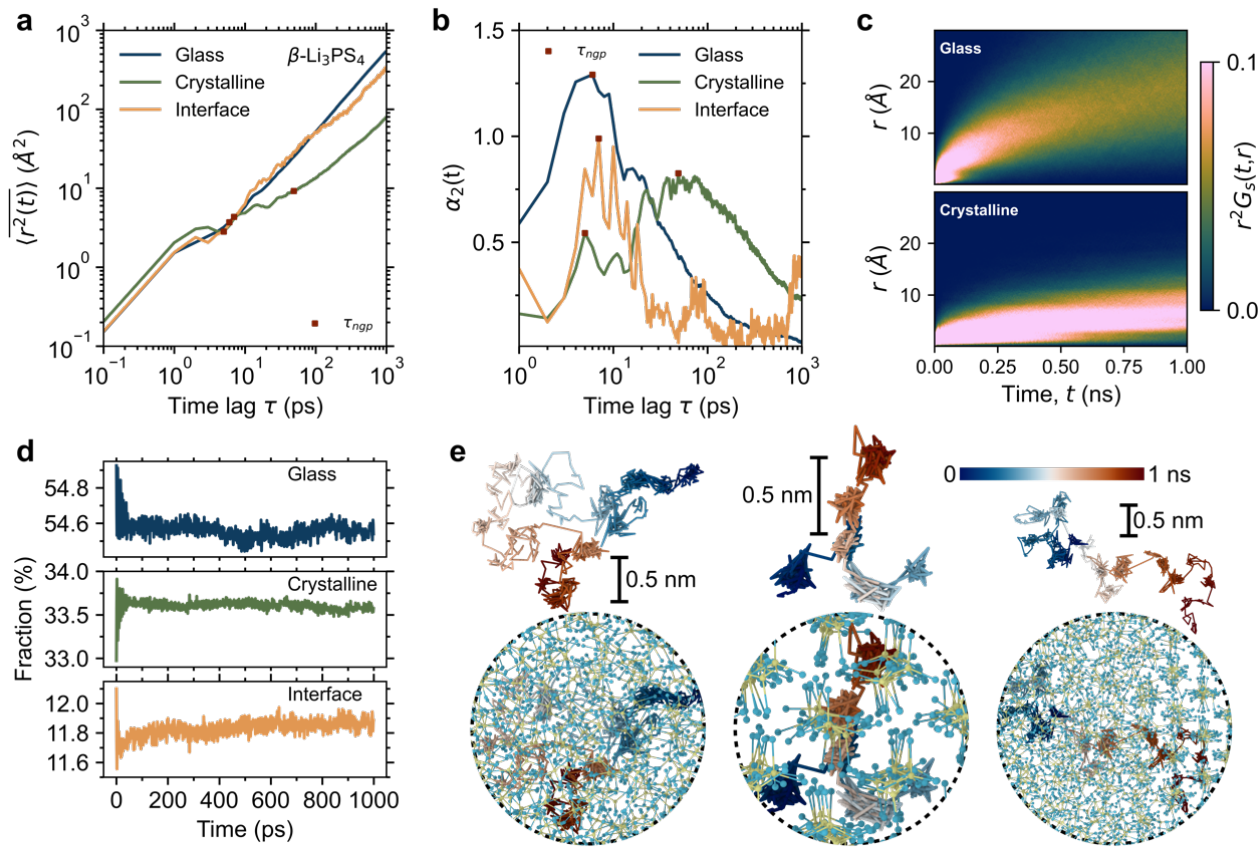

**Fig. 3 Diffusion dynamics from order to disorder. a** Time-averaged mean-squared displacement (MSD) and **b** non-Gaussian parameter (NGP) of lithium ions for glass-ceramic $Li_3PS_4$ system. **c** Self-part van Hove correlation function for glass-ceramic $Li_3PS_4$ system. **d** Time profiles of lithium-ion fraction within each phase of the glass-ceramic $Li_3PS_4$ system. **e** Atomic snapshots of lithium ions migration trajectories within glass (left), crystalline (center), and interfacial (right) phases of the glass-ceramic $Li_3PS_4$ system over a timespan of 1 ns.

**Fig. 3d** shows the time-concentration profiles of lithium ions in the various phases within the glass-ceramic $Li_3PS_4$. This analysis confirms that there is no significant enrichment of lithium ions in any phase during the diffusion process. **Fig. 3e** provides a visual representation of the trajectory of a single lithium-ion, highlighting the migration pathways of the lithium ions within the different phases of glass-ceramic $Li_3PS_4$. Specifically, in the disordered structure, ions navigate through the disordered potential-energy landscape[20], resulting in their meandering trajectories. Conversely, in the more ordered crystalline phase, distinct site-to-site hopping trajectories are apparent.

**Identifying ion conducting dynamics in both ordered and disordered structures.** In the case of crystals, ionic conductivity are related to the charge, concentration, and mobility of conducting ions[20]. The mechanism behind ion conduction can be explained through the hopping theory of conducting ions. In disordered structures, the lack of traditional coordination site and symmetric remote pathways[7], and the necessity for a percolating pathway of sites to minimize coordination changes[53,54], imply that the ion conduction is localized. That is, the ions are hopping between different sites, which possess different local environment. In both cases, the local coordination environment strongly impacts the ionic conduction. Here, we employ a classification-based machine learning approach, referred to as 'softness'[25–27], to establish a connection between the dynamics of hopping ions and the degree of structural order. Our previous work[55,56] has demonstrated the effectiveness of the softness method in capturing the local structural features of glassy electrolytes and establishing correlations between structure and the dynamics of conducting ions. To calculate softness, we first analyze the static structures and corresponding rearrangements of each lithium ion at 300 K. Subsequently, we employ logistic regression to establish a hyperplane for distinguishing 'mobile' from 'immobile' lithium ions, thereby determining their mobility characteristics, with softness ($S$) being defined as the distance to the feature space hyperplane. The mobility is analyzed using the non-affine square displacement ($D^2_{\min}$) for each lithium ion, with the sum of rearrangements for each lithium ion ($D_{\text{cum}}$) being used to quantify the extent of atomic rearrangements. For more information, we refer to the ***Methods*** section.

We plot the distribution of lithium softness $S$ values for β-$Li_3PS_4$, glassy $Li_3PS_4$ and glass-ceramic $Li_3PS_4$ in **Fig. 4a**. Positive values of $S$ correspond to mobility, whereas negative values signify immobility. The increase in disorder degree indeed shifts the softness distribution towards greater mobility, consistent with the observations reported above. The glass-ceramic $Li_3PS_4$ exhibits a bimodal distribution of $S$, with each peak corresponding to the softness distribution of immobile crystalline and mobile glassy phases, respectively, as illustrated in **Fig. S11**. This result is in line with the lower lithium-ion diffusion capability observed in crystalline phases compared to the glassy phases within glass-ceramic $Li_3PS_4$.

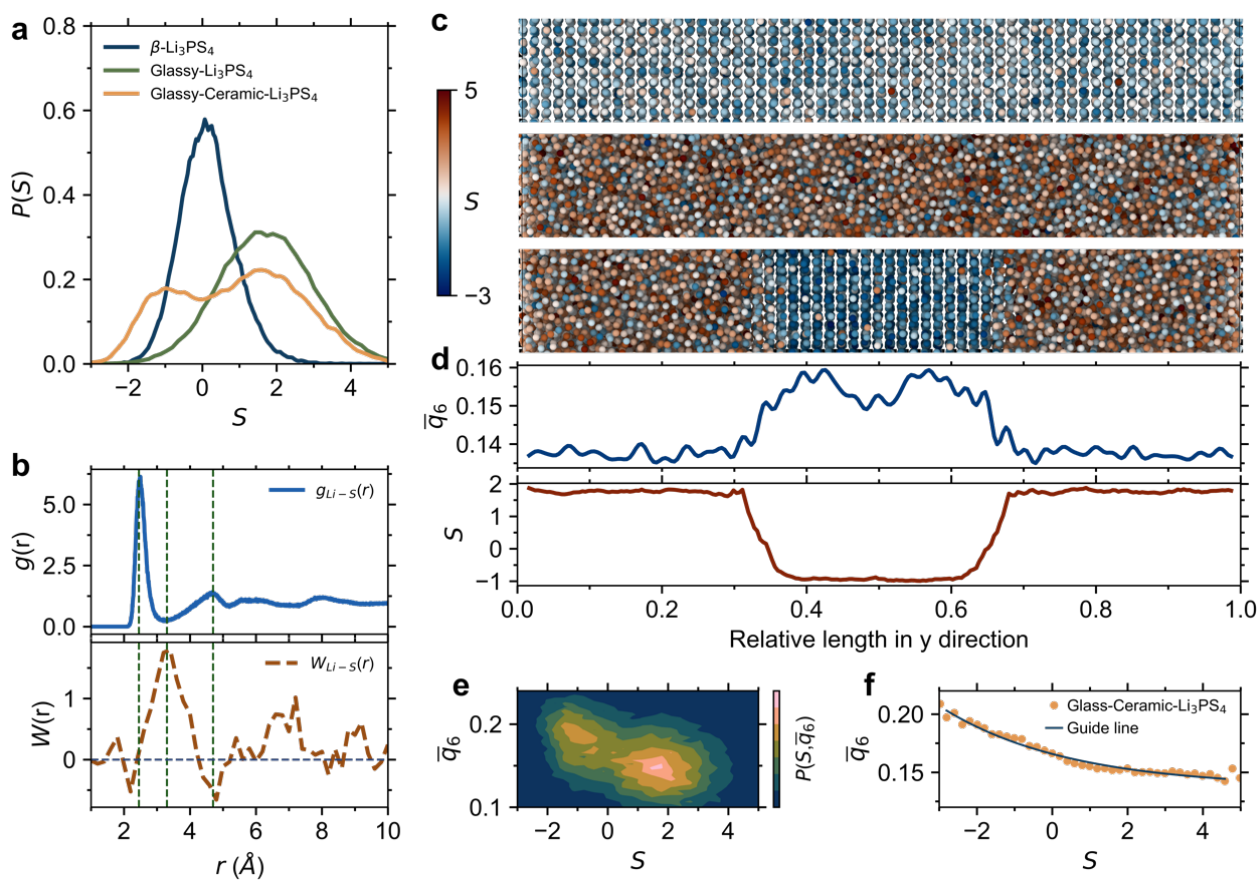

**Fig. 4 Machine-learning classified softness parameter. a** Distribution of lithium softness $S$ in β-, glassy-, and glass-ceramic $Li_3PS_4$ systems. **B** Radial distribution function (top panel) and weight function (bottom panel) of the Li-S pair for glass-ceramic $Li_3PS_4$ system. **c** Atomic snapshots of softness distribution in β-, glassy-, and glass-ceramic $Li_3PS_4$ systems. **d** Profiles of the averaged Steinhardt order parameter $\bar{q}_6$ and lithium softness $S$ along the $y$-direction in the glass-ceramic $Li_3PS_4$ configuration. **e** Density distribution of lithium softness $S$ and the averaged Steinhardt order parameter $\bar{q}_6$. **f** Correlation between the lithium softness $S$ value and $\bar{q}_6$.

The classification of lithium-ion mobility relies solely on the radial structure functions of Li-Li, Li-P, and Li-S pairs, achieving an accuracy of approximately 80%. Among these pairs, the $g(r)$ of Li-S for glass-ceramic $Li_3PS_4$ is the key correlation function to reflect the local structural environment of lithium ions, and hence dominates overall mobility. This aspect is illustrated in **Fig. 4b**, where $W(r)$ corresponds to the importance of each feature of $g(r)$. At the first peak of $g(r)$, the Li-S distance is at the equilibrium position, making it difficult for lithium ions to undergo significant rearrangement. As the distance increases, the Li-S separation starts to deviate from this equilibrium position. The increase in $W(r)$ indicates that the lithium ions become more likely to rearrange, reflecting a higher degree of 'softness'. In addition, we can observe positive values of $W(r)$ within the range of 6-8 Å, indicating that the MRO structure also influences the lithium-ion mobility. By coloring the lithium atoms based on their $S$ values, we can visualize the spatial softness distribution within the $Li_3PS_4$ structure (**Fig. 4c**), revealing a strong correlation between high $S$ values and structural disorder.

In the case of the glass-ceramic $Li_3PS_4$ electrolyte, we calculate the evolution of $S$ values within their structural landscape as the degree of disorder varies. To this end, we calculate a local bond-order parameter, known as the Steinhardt order parameter[57]. In detail, we employ the global average $\bar{q}_6$ parameter to characterize the structural order,[58] which involves the first and the second shell. **Fig. 4d** shows the calculate profile, demonstrating a strong correlation between $\bar{q}_6$ and $S$ values as the disorder-to-order transition occurs within the glass-ceramic $Li_3PS_4$ structure, with an abrupt transition at the interface. **Fig. 4e** presents the density distribution of $\bar{q}_6$ and softness values, with density values color-coded according to the magnitude of softness. High-softness regions are highlighted as areas with lower $\bar{q}_6$ values, while ions with low $S$ exhibit relatively higher $\bar{q}_6$ values. Therefore, we conclude that there exists a strong correlation between the structural order and the $S$ parameter, as quantified using the curve in **Fig. 4f**. This negative correlation between the $S$ values and the local structural order corroborates the enhanced dynamical properties of hopping ions in disordered environments. This approach is versatile and applicable not only to glassy and crystalline β-$Li_3PS_4$ electrolytes, as demonstrated in **Fig. 4**, but also to $Li_3PS_4$ systems with varying crystal contents and different crystal orientations, as shown in **Fig. S12** (cell parameters of β-$Li_3PS_4$ are shown in **Fig. S13**). This means that we can use this approach to predict the dynamic characteristics of the different phases solely based on their static structures, effectively distinguishing various phases with different degrees of softness.

**Discussion.** We have highlighted the enhancing effect of structural disorder on the dynamics of hopping ions and the synergistic role of disordered glass and interface phases in promoting ion conductivity. This holds significant implications for the conduction mechanisms of ionic species in solid-state electrolytes. Our work indicates that the disorder-driven diffusion dynamics suppresses the temperature impact of ionic conductivity, resulting in higher room-temperature conductivity in more disordered solid electrolytes compared to crystalline counterparts. Glass-ceramics also exhibit excellent ionic conductivity due to the dynamic interplay of their disordered glass phases and disordered interfaces. Furthermore, we have employed classification-based machine learning to gain insights into the mobility enhancement driven by disorder. This approach, linking dynamical characteristics with local structural information and atomic rearrangements, holds promise in unraveling ion transport mechanisms and discovering potential solid electrolytes for the development of high-performance all-solid-state lithium-ion batteries. Our study provides valuable insights into the intricate ion conduction mechanisms within complex disordered structures.

**References**
(1) Armand, M.; Tarascon, J.-M. Building Better Batteries. *Nature* **2008**, *451* (7179), 652–657. https://doi.org/10.1038/451652a.
(2) Scrosati, B.; Hassoun, J.; Sun, Y.-K. Lithium-Ion Batteries. A Look into the Future. *Energy Environ. Sci.* **2011**, *4* (9), 3287–3295. https://doi.org/10.1039/C1EE01388B.
(3) Chu, S.; Majumdar, A. Opportunities and Challenges for a Sustainable Energy Future. *Nature* **2012**, *488* (7411), 294–303. https://doi.org/10.1038/nature11475.
(4) Goodenough, J. B.; Park, K.-S. The Li-Ion Rechargeable Battery: A Perspective. *J. Am. Chem. Soc.* **2013**, *135* (4), 1167–1176. https://doi.org/10.1021/ja3091438.
(5) Janek, J.; Zeier, W. G. A Solid Future for Battery Development. *Nat Energy* **2016**, *1* (9), 16141. https://doi.org/10.1038/nenergy.2016.141.

(6) Liu, J.; Yuan, H.; Liu, H.; Zhao, C.; Lu, Y.; Cheng, X.; Huang, J.; Zhang, Q. Unlocking the Failure Mechanism of Solid State Lithium Metal Batteries. *Advanced Energy Materials* **2022**, *12* (4), 2100748. https://doi.org/10.1002/aenm.202100748.
(7) Famprikis, T.; Canepa, P.; Dawson, J. A.; Islam, M. S.; Masquelier, C. Fundamentals of Inorganic Solid-State Electrolytes for Batteries. *Nat. Mater.* **2019**, *18* (12), 1278–1291. https://doi.org/10.1038/s41563-019-0431-3.
(8) Inoue, T.; Mukai, K. Are All-Solid-State Lithium-Ion Batteries Really Safe?–Verification by Differential Scanning Calorimetry with an All-Inclusive Microcell. *ACS Appl. Mater. Interfaces* **2017**, *9* (2), 1507–1515. https://doi.org/10.1021/acsami.6b13224.
(9) Yan, J.; Zhu, D.; Ye, H.; Sun, H.; Zhang, X.; Yao, J.; Chen, J.; Geng, L.; Su, Y.; Zhang, P.; Dai, Q.; Wang, Z.; Wang, J.; Zhao, J.; Rong, Z.; Li, H.; Guo, B.; Ichikawa, S.; Gao, D.; Zhang, L.; Huang, J.; Tang, Y. Atomic-Scale Cryo-TEM Studies of the Thermal Runaway Mechanism of $Li_{1.3}Al_{0.3}Ti_{1.7}P_3O_{12}$ Solid Electrolyte. *ACS Energy Lett.* **2022**, *7* (11), 3855–3863. https://doi.org/10.1021/acsenergylett.2c01981.
(10) Das, A.; Sahu, S.; Mohapatra, M.; Verma, S.; Bhattacharyya, A. J.; Basu, S. Lithium-Ion Conductive Glass-Ceramic Electrolytes Enable Safe and Practical Li Batteries. *Materials Today Energy* **2022**, *29*, 101118. https://doi.org/10.1016/j.mtener.2022.101118.
(11) Chi, X.; Li, M.; Di, J.; Bai, P.; Song, L.; Wang, X.; Li, F.; Liang, S.; Xu, J.; Yu, J. A Highly Stable and Flexible Zeolite Electrolyte Solid-State Li–Air Battery. *Nature* **2021**, *592* (7855), 551–557. https://doi.org/10.1038/s41586-021-03410-9.
(12) Ravaine, D. Glasses as Solid Electrolytes. *Journal of Non-Crystalline Solids* **1980**, *38–39*, 353–358. https://doi.org/10.1016/0022-3093(80)90444-5.
(13) Zhang, Z.; Shao, Y.; Lotsch, B.; Hu, Y.-S.; Li, H.; Janek, J.; Nazar, L. F.; Nan, C.-W.; Maier, J.; Armand, M.; Chen, L. New Horizons for Inorganic Solid State Ion Conductors. *Energy Environ. Sci.* **2018**, *11* (8), 1945–1976. https://doi.org/10.1039/C8EE01053F.
(14) Choudhury, S.; Stalin, S.; Deng, Y.; Archer, L. A. Soft Colloidal Glasses as Solid-State Electrolytes. *Chem. Mater.* **2018**, *30* (17), 5996–6004. https://doi.org/10.1021/acs.chemmater.8b02227.
(15) Tatsumisago, M.; Hama, S.; Hayashi, A.; Morimoto, H.; Minami, T. New Lithium Ion Conducting Glass-Ceramics Prepared from Mechanochemical $Li_2S–P_2S_5$ Glasses. *Solid State Ionics* **2002**, *154–155*, 635–640. https://doi.org/10.1016/S0167-2738(02)00509-X.
(16) Mizuno, F.; Hayashi, A.; Tadanaga, K.; Tatsumisago, M. High Lithium Ion Conducting Glass-Ceramics in the System $Li_2S–P_2S_5$. *Solid State Ionics* **2006**, *177* (26–32), 2721–2725. https://doi.org/10.1016/j.ssi.2006.04.017.
(17) Seino, Y.; Ota, T.; Takada, K.; Hayashi, A.; Tatsumisago, M. A Sulphide Lithium Super Ion Conductor Is Superior to Liquid Ion Conductors for Use in Rechargeable Batteries. *Energy Environ. Sci.* **2014**, *7* (2), 627–631. https://doi.org/10.1039/C3EE41655K.
(18) Garcia-Mendez, R.; Smith, J. G.; Neuefeind, J. C.; Siegel, D. J.; Sakamoto, J. Correlating Macro and Atomic Structure with Elastic Properties and Ionic Transport of Glassy $Li_2S$-$P_2S_5$ (LPS) Solid Electrolyte for Solid-State Li Metal Batteries. *Adv. Energy Mater.* **2020**, *10* (19), 2000335. https://doi.org/10.1002/aenm.202000335.
(19) Chandra, A.; Bhatt, A.; Chandra, A. Ion Conduction in Superionic Glassy Electrolytes: An Overview. *Journal of Materials Science & Technology* **2013**, *29* (3), 193–208.

https://doi.org/10.1016/j.jmst.2013.01.005.
(20) Dyre, J. C.; Maass, P.; Roling, B.; Sidebottom, D. L. Fundamental Questions Relating to Ion Conduction in Disordered Solids. *Rep. Prog. Phys.* **2009**, *72* (4), 046501. https://doi.org/10.1088/0034-4885/72/4/046501.
(21) Bunde, A.; Funke, K.; Ingram, M. D. Ionic Glasses: History and Challenges. *Solid State Ionics* **1998**, *105* (1), 1–13. https://doi.org/10.1016/S0167-2738(97)00444-X.
(22) Ariga, S.; Ohkubo, T.; Urata, S.; Imamura, Y.; Taniguchi, T. A New Universal Force-Field for the $Li_2S$–$P_2S_5$ System. *Phys. Chem. Chem. Phys.* **2022**, *24* (4), 2567–2581. https://doi.org/10.1039/D1CP05393K.
(23) Kim, J.-S.; Jung, W. D.; Son, J.-W.; Lee, J.-H.; Kim, B.-K.; Chung, K.-Y.; Jung, H.-G.; Kim, H. Atomistic Assessments of Lithium-Ion Conduction Behavior in Glass–Ceramic Lithium Thiophosphates. *ACS Appl. Mater. Interfaces* **2019**, *11* (1), 13–18. https://doi.org/10.1021/acsami.8b17524.
(24) Wang, H.; Zhang, L.; Han, J.; E, W. DeePMD-Kit: A Deep Learning Package for Many-Body Potential Energy Representation and Molecular Dynamics. *Computer Physics Communications* **2018**, *228*, 178–184. https://doi.org/10.1016/j.cpc.2018.03.016.
(25) Cubuk, E. D.; Schoenholz, S. S.; Rieser, J. M.; Malone, B. D.; Rottler, J.; Durian, D. J.; Kaxiras, E.; Liu, A. J. Identifying Structural Flow Defects in Disordered Solids Using Machine-Learning Methods. *Phys. Rev. Lett.* **2015**, *114* (10), 108001. https://doi.org/10.1103/PhysRevLett.114.108001.
(26) Schoenholz, S. S.; Cubuk, E. D.; Sussman, D. M.; Kaxiras, E.; Liu, A. J. A Structural Approach to Relaxation in Glassy Liquids. *Nature Phys* **2016**, *12* (5), 469–471. https://doi.org/10.1038/nphys3644.
(27) Cubuk, E. D.; Ivancic, R. J. S.; Schoenholz, S. S.; Strickland, D. J.; Basu, A.; Davidson, Z. S.; Fontaine, J.; Hor, J. L.; Huang, Y.-R.; Jiang, Y.; Keim, N. C.; Koshigan, K. D.; Lefever, J. A.; Liu, T.; Ma, X.-G.; Magagnosc, D. J.; Morrow, E.; Ortiz, C. P.; Rieser, J. M.; Shavit, A.; Still, T.; Xu, Y.; Zhang, Y.; Nordstrom, K. N.; Arratia, P. E.; Carpick, R. W.; Durian, D. J.; Fakhraai, Z.; Jerolmack, D. J.; Lee, D.; Li, J.; Riggleman, R.; Turner, K. T.; Yodh, A. G.; Gianola, D. S.; Liu, A. J. Structure-Property Relationships from Universal Signatures of Plasticity in Disordered Solids. *Science* **2017**, *358* (6366), 1033–1037. https://doi.org/10.1126/science.aai8830.
(28) Wright, A. C. The Comparison of Molecular Dynamics Simulations with Diffraction Experiments. *Journal of Non-Crystalline Solids* **1993**, *159* (3), 264–268. https://doi.org/10.1016/0022-3093(93)90232-M.
(29) Homma, K.; Yonemura, M.; Kobayashi, T.; Nagao, M.; Hirayama, M.; Kanno, R. Crystal Structure and Phase Transitions of the Lithium Ionic Conductor $Li_3PS_4$. *Solid State Ionics* **2011**, *182* (1), 53–58. https://doi.org/10.1016/j.ssi.2010.10.001.
(30) Kaup, K.; Zhou, L.; Huq, A.; Nazar, L. F. Impact of the Li Substructure on the Diffusion Pathways in Alpha and Beta $Li_3PS_4$: An in Situ High Temperature Neutron Diffraction Study. *J. Mater. Chem. A* **2020**, *8* (25), 12446–12456. https://doi.org/10.1039/D0TA02805C.
(31) Stegmaier, S.; Schierholz, R.; Povstugar, I.; Barthel, J.; Rittmeyer, S. P.; Yu, S.; Wengert, S.; Rostami, S.; Kungl, H.; Reuter, K.; Eichel, R.; Scheurer, C. Nano-Scale Complexions Facilitate Li Dendrite-Free Operation in LATP Solid-State Electrolyte. *Adv. Energy Mater.* **2021**, *11* (26), 2100707. https://doi.org/10.1002/aenm.202100707.

(32) Dietrich, C.; Weber, D. A.; Sedlmaier, S. J.; Indris, S.; Culver, S. P.; Walter, D.; Janek, J.; Zeier, W. G. Lithium Ion Conductivity in $Li_2S–P_2S_5$ Glasses – Building Units and Local Structure Evolution during the Crystallization of Superionic Conductors $Li_3PS_4$, $Li_7P_3S_{11}$ and $Li_4P_2S_7$. *J. Mater. Chem. A* **2017**, *5* (34), 18111–18119. https://doi.org/10.1039/C7TA06067J.

(33) Self, E. C.; Chien, P.-H.; O'Donnell, L. F.; Morales, D.; Liu, J.; Brahmbhatt, T.; Greenbaum, S.; Nanda, J. Investigation of Glass-Ceramic Lithium Thiophosphate Solid Electrolytes Using NMR and Neutron Scattering. *Materials Today Physics* **2021**, *21*, 100478. https://doi.org/10.1016/j.mtphys.2021.100478.

(34) Sadowski, M.; Albe, K. Computational Study of Crystalline and Glassy Lithium Thiophosphates: Structure, Thermodynamic Stability and Transport Properties. *Journal of Power Sources* **2020**, *478*, 229041. https://doi.org/10.1016/j.jpowsour.2020.229041.

(35) Guo, H.; Wang, Q.; Urban, A.; Artrith, N. Artificial Intelligence-Aided Mapping of the Structure–Composition–Conductivity Relationships of Glass–Ceramic Lithium Thiophosphate Electrolytes. *Chem. Mater.* **2022**. https://doi.org/10.1021/acs.chemmater.2c00267.

(36) Mirmira, P.; Zheng, J.; Ma, P.; Amanchukwu, C. V. Importance of Multimodal Characterization and Influence of Residual $Li_2S$ Impurity in Amorphous $Li_3PS_4$ Inorganic Electrolytes. *J. Mater. Chem. A* **2021**, *9* (35), 19637–19648. https://doi.org/10.1039/D1TA02754A.

(37) Liu, Z.; Fu, W.; Payzant, E. A.; Yu, X.; Wu, Z.; Dudney, N. J.; Kiggans, J.; Hong, K.; Rondinone, A. J.; Liang, C. Anomalous High Ionic Conductivity of Nanoporous β-$Li_3PS_4$. *J. Am. Chem. Soc.* **2013**, *135* (3), 975–978. https://doi.org/10.1021/ja3110895.

(38) Tachez, M.; Malugani, J.-P.; Mercier, R.; Robert, G. Ionic Conductivity of and Phase Transition in Lithium Thiophosphate $Li_3PS_4$. *Solid State Ionics* **1984**, *14* (3), 181–185. https://doi.org/10.1016/0167-2738(84)90097-3.

(39) Yu, X.; Bates, J. B.; Jellison, G. E.; Hart, F. X. A Stable Thin-Film Lithium Electrolyte: Lithium Phosphorus Oxynitride. *J. Electrochem. Soc.* **1997**, *144* (2), 524. https://doi.org/10.1149/1.1837443.

(40) Kamaya, N.; Homma, K.; Yamakawa, Y.; Hirayama, M.; Kanno, R.; Yonemura, M.; Kamiyama, T.; Kato, Y.; Hama, S.; Kawamoto, K.; Mitsui, A. A Lithium Superionic Conductor. *Nature Mater* **2011**, *10* (9), 682–686. https://doi.org/10.1038/nmat3066.

(41) Buschmann, H.; Dölle, J.; Berendts, S.; Kuhn, A.; Bottke, P.; Wilkening, M.; Heitjans, P.; Senyshyn, A.; Ehrenberg, H.; Lotnyk, A.; Duppel, V.; Kienle, L.; Janek, J. Structure and Dynamics of the Fast Lithium Ion Conductor "$Li_7La_3Zr_2O_{12}$." *Physical Chemistry Chemical Physics* **2011**, *13* (43), 19378–19392. https://doi.org/10.1039/C1CP22108F.

(42) Francisco, B. E.; Stoldt, C. R.; M'Peko, J.-C. Lithium-Ion Trapping from Local Structural Distortions in Sodium Super Ionic Conductor (NASICON) Electrolytes. *Chem. Mater.* **2014**, *26* (16), 4741–4749. https://doi.org/10.1021/cm5013872.

(43) Edman, L.; Ferry, A.; Doeff, M. M. Slow Recrystallization in the Polymer Electrolyte System Poly(Ethylene Oxide)$_n$–$LiN(CF_3SO_2)_2$. *Journal of Materials Research* **2000**, *15* (9), 1950–1954. https://doi.org/10.1557/JMR.2000.0281.

(44) Stallworth, P. E.; Fontanella, J. J.; Wintersgill, M. C.; Scheidler, C. D.; Immel, J. J.; Greenbaum, S. G.; Gozdz, A. S. NMR, DSC and High Pressure Electrical Conductivity Studies of Liquid and Hybrid Electrolytes. *Journal of Power Sources* **1999**, *81–82*, 739–747. https://doi.org/10.1016/S0378-7753(99)00144-5.

(45) Poletayev, A. D.; Dawson, J. A.; Islam, M. S.; Lindenberg, A. M. Defect-Driven Anomalous Transport in Fast-Ion Conducting Solid Electrolytes. *Nat. Mater.* **2022**, *21* (9), 1066–1073. https://doi.org/10.1038/s41563-022-01316-z.

(46) Song, S.; Park, S. J.; Kim, M.; Kim, J. S.; Sung, B. J.; Lee, S.; Kim, J.-H.; Sung, J. Transport Dynamics of Complex Fluids. *Proceedings of the National Academy of Sciences* **2019**, *116* (26), 12733–12742. https://doi.org/10.1073/pnas.1900239116.

(47) Forrester, F. N.; Quirk, J. A.; Famprikis, T.; Dawson, J. A. Disentangling Cation and Anion Dynamics in $Li_3PS_4$ Solid Electrolytes. *Chem. Mater.* **2022**, *34* (23), 10561–10571. https://doi.org/10.1021/acs.chemmater.2c02637.

(48) Hayashi, A.; Hama, S.; Minami, T.; Tatsumisago, M. Formation of Superionic Crystals from Mechanically Milled $Li_2S$–$P_2S_5$ Glasses. *Electrochemistry Communications* **2003**, *5* (2), 111–114. https://doi.org/10.1016/S1388-2481(02)00555-6.

(49) Tsukasaki, H.; Mori, S.; Shiotani, S.; Yamamura, H. Ionic Conductivity and Crystallization Process in the $Li_2S$–$P_2S_5$ Glass Electrolyte. *Solid State Ionics* **2018**, *317*, 122–126. https://doi.org/10.1016/j.ssi.2018.01.010.

(50) Eom, M.; Kim, J.; Noh, S.; Shin, D. Crystallization Kinetics of $Li_2S$–$P_2S_5$ Solid Electrolyte and Its Effect on Electrochemical Performance. *Journal of Power Sources* **2015**, *284*, 44–48. https://doi.org/10.1016/j.jpowsour.2015.02.141.

(51) Uchida, K.; Ohkubo, T.; Utsuno, F.; Yazawa, K. Modified $Li_7P_3S_{11}$ Glass-Ceramic Electrolyte and Its Characterization. *ACS Appl. Mater. Interfaces* **2021**, *13* (31), 37071–37081. https://doi.org/10.1021/acsami.1c08507.

(52) Mizuno, F.; Hayashi, A.; Tadanaga, K.; Tatsumisago, M. New, Highly Ion-Conductive Crystals Precipitated from $Li_2S$–$P_2S_5$ Glasses. *Advanced Materials* **2005**, *17* (7), 918–921. https://doi.org/10.1002/adma.200401286.

(53) Zeng, Y.; Ouyang, B.; Liu, J.; Byeon, Y.-W.; Cai, Z.; Miara, L. J.; Wang, Y.; Ceder, G. High-Entropy Mechanism to Boost Ionic Conductivity. *Science* **2022**, *378* (6626), 1320–1324. https://doi.org/10.1126/science.abq1346.

(54) Qiao, A.; Ren, J.; Tao, H.; Zhao, X.; Yue, Y. Percolative Channels for Superionic Conduction in an Amorphous Conductor. *J. Phys. Chem. Lett.* **2022**, *13* (45), 10507–10512. https://doi.org/10.1021/acs.jpclett.2c02776.

(55) Chen, Z.; Du, T.; Christensen, R.; Bauchy, M.; Smedskjaer, M. M. Deciphering How Anion Clusters Govern Lithium Conduction in Glassy Thiophosphate Electrolytes through Machine Learning. *ACS Energy Lett.* **2023**, 1969–1975. https://doi.org/10.1021/acsenergylett.3c00237.

(56) Du, T.; Chen, Z.; Liu, H.; Zhang, Q.; Bauchy, M.; Yue, Y.; Smedskjaer, M. M. Controlling Factor for Fracture Resistance and Ionic Conduction in Glassy Lithium Borophosphate Electrolytes. *Materials Today Energy* **2023**, *37*, 101390. https://doi.org/10.1016/j.mtener.2023.101390.

(57) Steinhardt, P. J.; Nelson, D. R.; Ronchetti, M. Bond-Orientational Order in Liquids and Glasses. *Phys. Rev. B* **1983**, *28* (2), 784–805. https://doi.org/10.1103/PhysRevB.28.784.

(58) Lechner, W.; Dellago, C. Accurate Determination of Crystal Structures Based on Averaged Local Bond Order Parameters. *The Journal of Chemical Physics* **2008**, *129* (11), 114707. https://doi.org/10.1063/1.2977970.

(59) Harris, C. R.; Millman, K. J.; van der Walt, S. J.; Gommers, R.; Virtanen, P.; Cournapeau, D.; Wieser, E.; Taylor, J.; Berg, S.; Smith, N. J.; Kern, R.; Picus, M.; Hoyer, S.; van Kerkwijk, M. H.;

Brett, M.; Haldane, A.; del Río, J. F.; Wiebe, M.; Peterson, P.; Gérard-Marchant, P.; Sheppard, K.; Reddy, T.; Weckesser, W.; Abbasi, H.; Gohlke, C.; Oliphant, T. E. Array Programming with NumPy. *Nature* **2020**, *585* (7825), 357–362. https://doi.org/10.1038/s41586-020-2649-2.
(60) McKinney, W. Data Structures for Statistical Computing in Python. *Proceedings of the 9th Python in Science Conference* **2010**, 56–61. https://doi.org/10.25080/Majora-92bf1922-00a.
(61) Virtanen, P.; Gommers, R.; Oliphant, T. E.; Haberland, M.; Reddy, T.; Cournapeau, D.; Burovski, E.; Peterson, P.; Weckesser, W.; Bright, J.; van der Walt, S. J.; Brett, M.; Wilson, J.; Millman, K. J.; Mayorov, N.; Nelson, A. R. J.; Jones, E.; Kern, R.; Larson, E.; Carey, C. J.; Polat, İ.; Feng, Y.; Moore, E. W.; VanderPlas, J.; Laxalde, D.; Perktold, J.; Cimrman, R.; Henriksen, I.; Quintero, E. A.; Harris, C. R.; Archibald, A. M.; Ribeiro, A. H.; Pedregosa, F.; van Mulbregt, P. SciPy 1.0: Fundamental Algorithms for Scientific Computing in Python. *Nat Methods* **2020**, *17* (3), 261–272. https://doi.org/10.1038/s41592-019-0686-2.
(62) Hunter, J. D. Matplotlib: A 2D Graphics Environment. *Computing in Science & Engineering* **2007**, *9* (03), 90–95. https://doi.org/10.1109/MCSE.2007.55.
(63) Stukowski, A. Visualization and Analysis of Atomistic Simulation Data with OVITO–the Open Visualization Tool. *Modelling Simul. Mater. Sci. Eng.* **2009**, *18* (1), 015012. https://doi.org/10.1088/0965-0393/18/1/015012.
(64) Kohn, W.; Sham, L. J. Self-Consistent Equations Including Exchange and Correlation Effects. *Phys. Rev.* **1965**, *140* (4A), A1133–A1138. https://doi.org/10.1103/PhysRev.140.A1133.
(65) Kühne, T. D.; Iannuzzi, M.; Del Ben, M.; Rybkin, V. V.; Seewald, P.; Stein, F.; Laino, T.; Khaliullin, R. Z.; Schütt, O.; Schiffmann, F.; Golze, D.; Wilhelm, J.; Chulkov, S.; Bani-Hashemian, M. H.; Weber, V.; Borštnik, U.; Taillefumier, M.; Jakobovits, A. S.; Lazzaro, A.; Pabst, H.; Müller, T.; Schade, R.; Guidon, M.; Andermatt, S.; Holmberg, N.; Schenter, G. K.; Hehn, A.; Bussy, A.; Belleflamme, F.; Tabacchi, G.; Glöß, A.; Lass, M.; Bethune, I.; Mundy, C. J.; Plessl, C.; Watkins, M.; VandeVondele, J.; Krack, M.; Hutter, J. CP2K: An Electronic Structure and Molecular Dynamics Software Package - Quickstep: Efficient and Accurate Electronic Structure Calculations. *The Journal of Chemical Physics* **2020**, *152* (19), 194103. https://doi.org/10.1063/5.0007045.
(66) VandeVondele, J.; Krack, M.; Mohamed, F.; Parrinello, M.; Chassaing, T.; Hutter, J. Quickstep: Fast and Accurate Density Functional Calculations Using a Mixed Gaussian and Plane Waves Approach. *Computer Physics Communications* **2005**, *167* (2), 103–128. https://doi.org/10.1016/j.cpc.2004.12.014.
(67) Nosé, S. A Molecular Dynamics Method for Simulations in the Canonical Ensemble. *Molecular Physics* **1984**, *52* (2), 255–268. https://doi.org/10.1080/00268978400101201.
(68) Perdew, J. P.; Burke, K.; Ernzerhof, M. Generalized Gradient Approximation Made Simple. *Phys. Rev. Lett.* **1996**, *77* (18), 3865–3868. https://doi.org/10.1103/PhysRevLett.77.3865.
(69) Grimme, S.; Antony, J.; Ehrlich, S.; Krieg, H. A Consistent and Accurate Ab Initio Parametrization of Density Functional Dispersion Correction (DFT-D) for the 94 Elements H-Pu. *The Journal of Chemical Physics* **2010**, *132* (15), 154104. https://doi.org/10.1063/1.3382344.
(70) Goedecker, S.; Teter, M.; Hutter, J. Separable Dual-Space Gaussian Pseudopotentials. *Phys. Rev. B* **1996**, *54* (3), 1703–1710. https://doi.org/10.1103/PhysRevB.54.1703.
(71) Zhang, Y.; Wang, H.; Chen, W.; Zeng, J.; Zhang, L.; Wang, H.; E, W. DP-GEN: A Concurrent Learning Platform for the Generation of Reliable Deep Learning Based Potential Energy Models. *Computer Physics Communications* **2020**, *253*, 107206. https://doi.org/10.1016/j.cpc.2020.107206.

(72) Plimpton, S. Fast Parallel Algorithms for Short-Range Molecular Dynamics. *Journal of Computational Physics* **1995**, *117* (1), 1–19. https://doi.org/10.1006/jcph.1995.1039.
(73) Faber, T. E.; Ziman, J. M. A Theory of the Electrical Properties of Liquid Metals. *The Philosophical Magazine: A Journal of Theoretical Experimental and Applied Physics* **1965**, *11* (109), 153–173. https://doi.org/10.1080/14786436508211931.
(74) Liu, H.; Xiao, S.; Tang, L.; Bao, E.; Li, E.; Yang, C.; Zhao, Z.; Sant, G.; Smedskjaer, M. M.; Guo, L.; Bauchy, M. Predicting the Early-Stage Creep Dynamics of Gels from Their Static Structure by Machine Learning. *Acta Materialia* **2021**, *210*, 116817. https://doi.org/10.1016/j.actamat.2021.116817.
(75) Sierou, A.; Brady, J. F. Shear-Induced Self-Diffusion in Non-Colloidal Suspensions. *Journal of Fluid Mechanics* **2004**, *506*, 285–314. https://doi.org/10.1017/S0022112004008651.

**Methods**

**General methods.** We employed the Python packages NumPy[59], Pandas[60] and SciPy[61] for data processing and calculations, and utilized Matplotlib[62] and Ovito[63] for graph generation and the creation of renderable structural visualizations.

**Ab initio simulations.** The initial datasets for training the MLIP were generated from the trajectories of *ab initio* molecular dynamics (AIMD) simulations. To increase the generalizability of the MLIP, series of systems containing Li, P, and S elements were covered in the datasets, including both the crystalline and disordered structures of elementary substances and compounds, such as Li, P, S, $Li_2P_2S_6$, β-$Li_3PS_4$, and γ-$Li_3PS_4$, as well as $xLi_2S$-$(100-x)P_2S_5$ ($x$ = 67, 70, 75, and 80) glasses. Detailed information on the included systems can be found in the Supporting **Table S1**. The AIMD calculations were carried out at the DFT level[64] using the Quickstep module of the CP2K package[65] with the hybrid Gaussian and plane wave method (GPW)[66]. To ensure computational accuracy, the basis functions were mapped onto a multi-grid system with the default number of four different grids with a plane-wave cutoff for the electronic density of 500 Ry and a relative cutoff of 50 Ry. The AIMD trajectories at 3000 K were obtained in the *NVT* ensemble with a timestep of 0.5 fs for 2.5 ps. The temperature selection of 3000 K enabled the sampling of the melting process within the short time scale, which can be used for simulating both the crystal and glass structure afterwards. The temperature was controlled using the Nosé–Hoover thermostat[67]. The exchange-correlation energy was calculated using the Perdew-Burke-Ernzerhof (PBE) approximation[68], and the dispersion interactions were handled by utilizing the empirical dispersion correction (D3) from Grimme[69]. The pseudopotential GTH-PBE combined with the corresponding basis sets were employed to describe the valence electrons of Li (DZVP-MOLOPT-SR-GTH), P (TZVP-MOLOPT-GTH), and S (TZVP-MOLOPT-GTH), respectively[70].

**MLIP training and validation.** The MLIP was developed using the DeePMD-kit[24] software combined with the active machine learning method implemented in the DP-GEN package[71]. The training dataset consisted of two parts: (1) the initial training data from the trajectories of AIMD simulations on various Li, P, and S-containing systems at 3000 K for 2.5 ps; and (2) the expanded dataset realized by single energy calculation of different $xLi_2S$-$(100-x)P_2S_5$ ($x$ = 67, 70, 75) glasses using the active machine learning method implemented in the DP-GEN package[71]. The detailed information of the two datasets can be found in the Supporting **Table S1** and **S2**, respectively. The network structure used for training the MLIP also consisted of two parts: (1) the embedding network with 3 layers of neurons (25, 50, 100); and (2) the fitting network with 3 layers of neurons (240, 240, 240). The local environment of the individual atoms was described using the descriptor containing both radial and angular information within a cutoff of 6.5 Å. The two training processes were adopted to increase the accuracy of the MLIP, i.e., the initial training and final training both included 6,000,000 steps of iterations. The energy, force, and virial terms were included in the loss function, which enabled the MLIP to work for both the structure and mechanical simulations. For the initial training, the learning rate dynamically changed from 1e-3 to 1e-9. The atomic interactions beyond 0.9 Å were treated using the ZBL repulsive interactions to avoid the collapse of atoms at high temperatures. The prefactors of the energy, force, and virial terms dynamically changed from 0.02 to 2, 1000 to 1, and 0.02 to 0.2, respectively. During the final training, the ZBL interaction was removed and the prefactors of all terms were set to 1 as the learning rate changed from 1e-5 to 1e-8. The performance of the MLIP model on different $Li_3PS_4$ phases is shown in **Fig. S14**, providing compelling evidence of its ability to accurately describe interatomic interactions.

**MD simulations.** MD simulations were performed using the Large-scale Atomic/Molecular Massively Parallel Simulator (LAMMPS)[72]. The neural network potential trained using the DeePMD method, as described above, was employed to accurately describe the interatomic potentials. The temperature and pressure were controlled using Nosé-Hoover[67] thermostat/barostat methods. β-$Li_3PS_4$ with the *Pnma* space group was annealed from 500 K and relaxed for 1 ns at 300 K using the *NPT* ensemble. The preparation of glassy $Li_3PS_4$ was done using melt-quenching by initially raising the system to 1500 K, holding it for 100 ps, and subsequently cooling it down to 300 K at a rate of 2.5K/ps before relaxation, all under the *NPT* ensemble. A time step of 0.5 fs was employed for precise and stable structural simulations throughout. The glass-ceramic $Li_3PS_4$ was made analogously to the glass, with a melt-quenching process under the *NVT* ensemble, during which the crystalline region was frozen (with all force components set to zero), followed by relaxation for glass and crystalline region under the *NPT* ensemble.

**Structural descriptors.** The partial radial distribution functions $g_{ij}(r)$ define the probability of finding a particle *j* at distance $r + \Delta r$ given that there is a particle *i*,

$$g_{ij}(r) = \frac{n_{ij}(r)}{4\pi r^2 \mathrm{d}r \rho_j}, \qquad (1)$$

where $n_{ij}$ is the number of *j*-type atoms found in a spherical shell of radius *r* and thickness $\Delta r$, with the centra *i*-type atom. $P_j$ is the number density of *j*-type atoms. The running coordination $n®$ is obtained by intergrating $g_{ij}(r)$ between $r_1$ and $r_2$ as,

$$n(r) = \int_{r_1}^{r_2} 4\pi r^2 \rho_j g_{ij}(r) \mathrm{d}r. \qquad (2)$$

The theoretical partial structure factor $S_{ij}(Q)$ was calculated using the Faber Ziman formalism[73] as,

$$S_{ij}(Q) - 1 = \rho \int_0^\infty 4\pi r^2 \big(g_{ij}(r) - 1\big) \frac{\sin Qr}{Qr} \mathrm{d}r, \qquad (3)$$

where *Q* is the scattering vector, and the X-ray scattering structure factor $S_x(Q)$ was calculated as,

$$S_X(Q) = \sum_i \sum_{j=i} c_i c_j f_i(Q) f_j(Q) \big(S_{ij}(Q) - 1\big). \quad (4)$$

Here, $c_i$ and $c_j$ are the atomic fractions, $f_i(Q)$ and $f_j(Q)$ refer to the $Q$-dependent X-ray scattering coefficients of type $I$ and $j$ atoms, respectively. By considering the coherent scattering length $\bar{I}$ and $\bar{b}_j$ of atoms, the neutron scattering structure factor $S_N(Q)$ was calculated as,

$$S_N(Q) = \sum_i \sum_{j=i} c_i c_j \bar{b}_i \bar{b}_j \big(S_{ij}(Q) - 1\big). \quad (5)$$

The degree of disorder was quantified using the function $F(Z)$. We followed a procedure similar to that in ref.[31], beginning with a 3D Gaussian density distribution of the atomic positions. After mapping these positions onto a 3D grid, we created density slabs along the $y$-axis of the simulation box with a width of $\Delta y$ and projected them into 2D. By summing the intensities, which were obtained through a 2D discrete Fourier Transform with the zero-frequency component shifted to the center, we calculated $F(Z)$, which was then normalized to its maximum value,

$$F(z) = \frac{\sum_{xz} I_{2D-FFT}(y)}{\sum_{xz} I_{max}}. \quad (6)$$

To distinguish between ordered crystalline phases and disordered glassy phases, we introduced an average local bond order parameter, or the average Steinhardt order parameter. Based on the spherical harmonic algorithm, a complex vector for a particle can be defined as $q_{lm}$ (**eq. 7**), where $N_b$ is the number of nearest neighbors for particle $i$, and $Y_{lm}$ represents spherical harmonics,

$$q_{lm}(i) = \frac{1}{N_b(i)} \sum_{j=1}^{N_b(i)} Y_{lm}(\mathbf{r}_{ij}). \quad (7)$$

The Steinhardt order parameters[57] are defined as,

$$q_l(i) = \sqrt{\frac{4\pi}{2l+1} \sum_{m=-l}^{l} |q_{lm}(i)|^2}. \quad (8)$$

Here, we performed a second averaging over the first shell of particles, implicitly incorporating information from the second shell[58], which was computed by replacing $q_{lm}(i)$ with $\bar{q}_{lm}(i)$. The average value of $q_{lm}(i)$ over all the $N_b$ neighbors $k$ of particle $i$, including particle $i$ itself, was calculated as,

$$\bar{q}_{lm}(i) = \frac{1}{N_b} \sum_{k=0}^{N_b} q_{lm}(k). \quad (9)$$

**Lithium transport dynamics.** The mean-squared displacement (MSD) and non-Gaussian parameter were calculated from long-time lag ($t$) trajectory as,

$$\overline{\langle r^2(t) \rangle} = \Delta r^2(t) = \langle (r(t) - r(0))^2 \rangle, \quad (10)$$

$$\alpha_2(t) = \frac{d\langle (r(t)-r(0))^4 \rangle}{(d+2)\langle (r(t)-r(0))^2 \rangle^2} - 1, \quad (11)$$

where the angular brackets denote an ensemble-averaged over the total conduction atoms, i.e., Li ions, and $d$ is the dimension of the simulation box, where $d = 3$ for all simulations. The self-part of the van Hove correlation function $G_s$ for Li-Li pair was calculated as,

$$G_s(r,t) = \frac{1}{4\pi r^2 N} \left\langle \sum_{i=1}^{N} \delta[r - |r_i(t_0) - r_i(t+t_0)|] \right\rangle_{t_0}, \quad (12)$$

where $G_s(r,t)$ characterizes the Li-Li pair distance $r$ at time $t$, and the quantity $r^2 G_s(r,t)$ describes the probability distribution of displacements.

The self-diffusion coefficient $D$ was estimated from the MSD and hence the activation energy $E_a$ was calculated by fitting an Arrhenius function as,

$$D = \frac{1}{2d} \lim_{t \to \infty} \frac{d\langle MSD \rangle}{dt}, \quad (13)$$

$$D = D_0 \exp(-E_a/k_B T), \quad (14)$$

where $k_B$ is the Boltzmann constant, $T$ is the temperature, and $D_0$ is the self-diffusion coefficient at an infinite temperature. Finally, by means of *Nernst-Einstein* equation using the elementary charge $e$, the ionic conductivity $\sigma$ was calculated,

$$\sigma = \frac{N}{V} \frac{(Ze)^2}{k_B T} D. \quad (15)$$

**Machine learning classification.** We calculated the 'softness' metric[25–27] based on classification-based machine learning, following the procedure outlined in refs.[55,56]. Unlike the original concept of softness, the computation of softness here relies on a logistic regression classifier instead of a support vector machine due to its higher classification accuracy and training efficiency[74,55,56]. Softness is defined as the distance to the feature space hyperplane, with radial order parameters chosen as the features for constructing the hyperplane. The hyperplane created by logistic regression can then be expressed as a function of each feature,

$$\sum_r W(r)G(i;r) - b = 0, \quad (16)$$

where the feature $G(i, r)$ represents the standardized radial order parameters, being a function of pairwise distance $r$. Here, $W(r)$ and $b$ are the weight coefficients and bias of the logistic regression model, respectively. The hyperplane is a linear combination of input features, allowing softness to be determined based on different features. In other words, the absolute value of $W(r)$ indicates the importance of the corresponding feature $G(i, r)$, with positive and negative signs signifying that an increase in the value of $G(i, r)$ will, respectively, increase or decrease the softness value.

The logistic regression model's output results were chosen based on the $D_{cum}$ metric for each lithium atom to analyze ion conduction behavior. The $D_{cum}$ is the sum of square-root of the incremental non-affine squared displacement $D_{min}^2$, which has been widely used to describe atomic rearrangement processes[25,75]. We optimized the intervals ($dr$) and cutoff radius ($R_{cutoff}$) of the radial order parameters based on their classification accuracy, as illustrated in **Fig. S15**. By employing $dr$ of 0.2 Å and $R_{cutoff}$ of 10 Å, we established structural features that yielded prediction accuracy exceeding 0.78 for the test set. Regularization parameters $C$ and the threshold for $D_{cum}$ were also determined based on classification accuracy, as shown in **Fig. S16**. Both the training and test sets achieved accuracies exceeding 80%. With these adjustments, we have constructed a framework capable of predicting the ionic conductivity dynamics based on the softness properties.

**Data availability**

The templates to generate glass and glass-ceramic structures and to run simulations, and neuron-network potential file are available at https://github.com/OxideGlassGroupAAU/LiPS. Source data are provided with this paper.

**Code availability**

LAMMPS and DeePMD-kit are free and open-source codes available at https://lammps.sandia.gov and https://github.com/deepmodeling/deepmd-kit, respectively.

**Acknowledgments**
This work was supported by grants from China Scholarship Council (202106880010) and the European Union (ERC, NewGLASS, 101044664). Views and opinions expressed are, however, those of the authors only and do not necessarily reflect those of the European Union or the European Research Council. Neither the European Union nor the granting authority can be held responsible for them. We also acknowledge the computational resources supplied by EuroHPC Joint Undertaking with access to Vega at IZUM, Slovenia (EHPC-REG-2022R02-224) and Aalborg University (CLAAUDIA).

**Author contributions**
Z.C., T.D. and M.M.S. conceived the study and planned the computational work. Z.C. performed the molecular dynamics simulations, structural and dynamical analyses, and machine-learning calculations with input and assistance from T.D., N.M.A.K., Y.Y. and M.M.S. T.D. trained the neural-network potential. M.M.S. supervised the study. All authors participated in discussing the data and contributed to the writing of the manuscript.

**Competing interests**
The authors declare no competing interests.

# SUPPORTING INFORMATION for

# Disorder-induced enhancement of lithium-ion transport in solid-state electrolytes

Zhimin Chen[1], Tao Du[1,*], N. M. Anoop Krishnan[2], Yuanzheng Yue[1], Morten M. Smedskjaer[1,*]

[1] *Department of Chemistry and Bioscience, Aalborg University, Aalborg East 9220, Denmark*
[2] *Department of Civil Engineering, Indian Institute of Technology Delhi, New Delhi 110016, India*
\* *Corresponding authors. E-mail:* [taod@bio.aau.dk](taod@bio.aau.dk) *(T.D.),* [mos@bio.aau.dk](mos@bio.aau.dk) *(M.M.S.)*

**Supporting Figures**

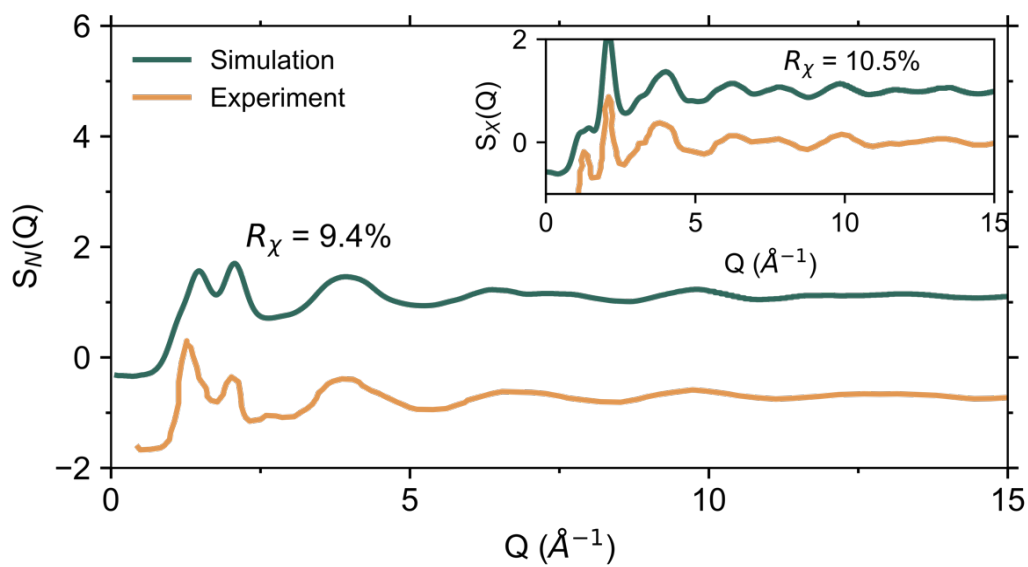

**Fig. S1.** Neutron structure factor, $S_N(Q)$, comparison from MD simulations (using the present MLIP) and neutron scattering experiments for glassy $Li_3PS_4$. The inset shows the X-ray structure factor, $S_X(Q)$, comparison with X-ray scattering experiments.

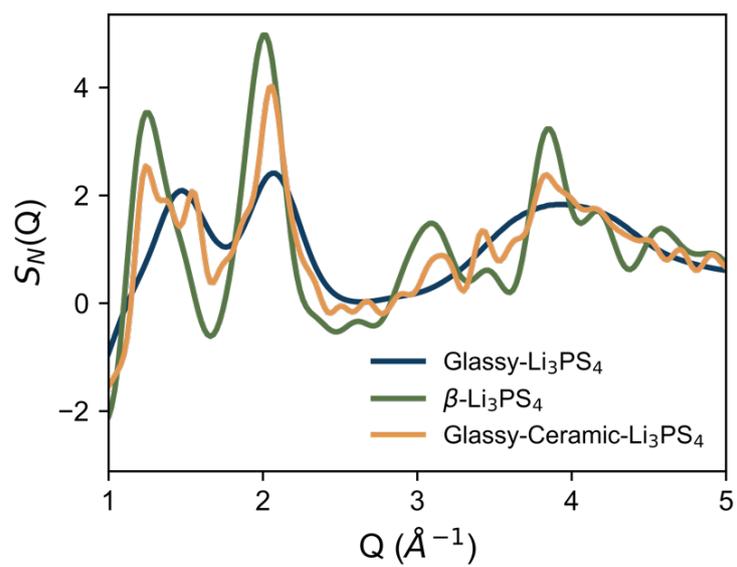

**Fig. S2.** Neutron structure factor $S_N(Q)$ of simulated glassy, β-, and glass-ceramic $Li_3PS_4$ electrolytes using the MLP.

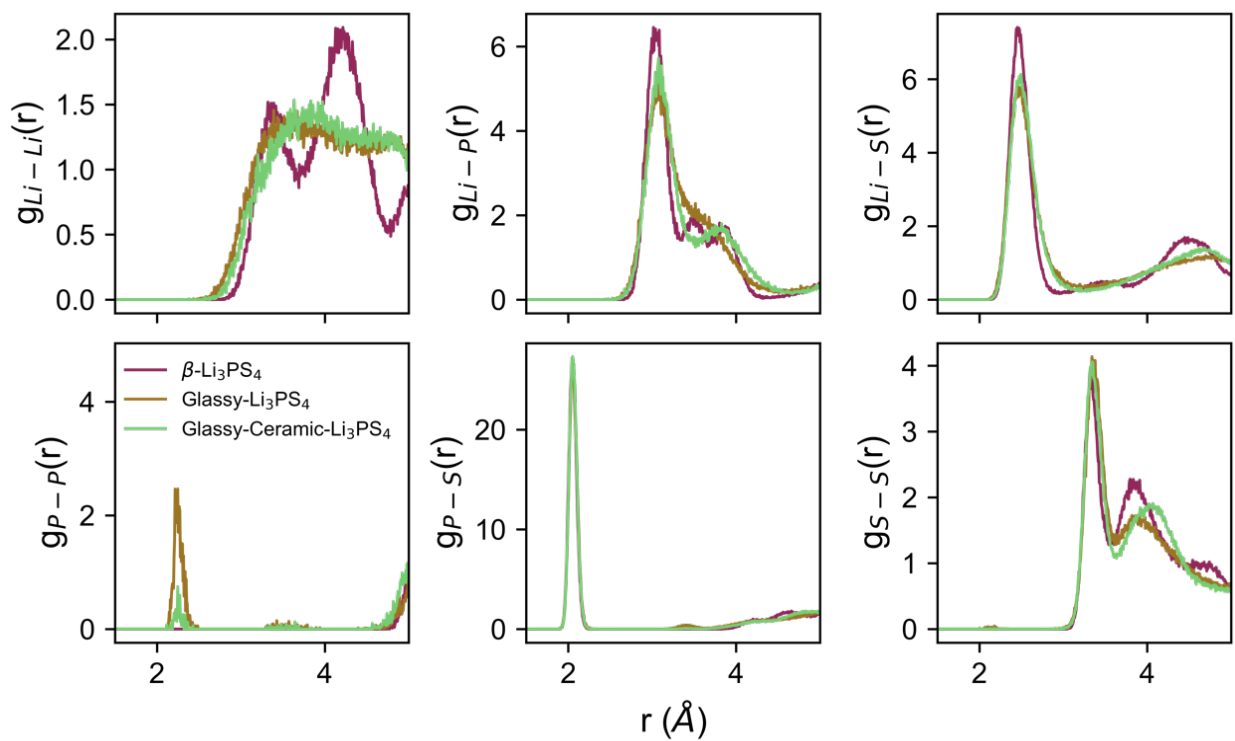

**Fig. S3.** Partial radial distribution functions $g_{ij}(r)$ of all atomic pairs in the simulated glassy, β-, and glass-ceramic $Li_3PS_4$ electrolytes using the MLIP.

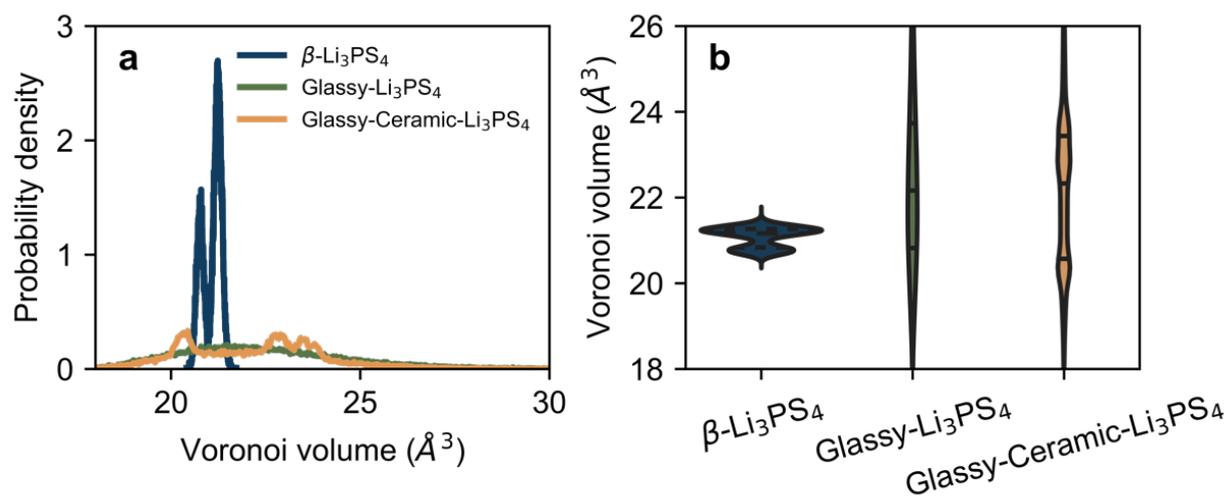

**Fig. S4.** Voronoi volume of lithium atoms in simulated Li$_3$PS$_4$ electrolytes using the MLIP potential: (a) density distribution and (b) violin plot.

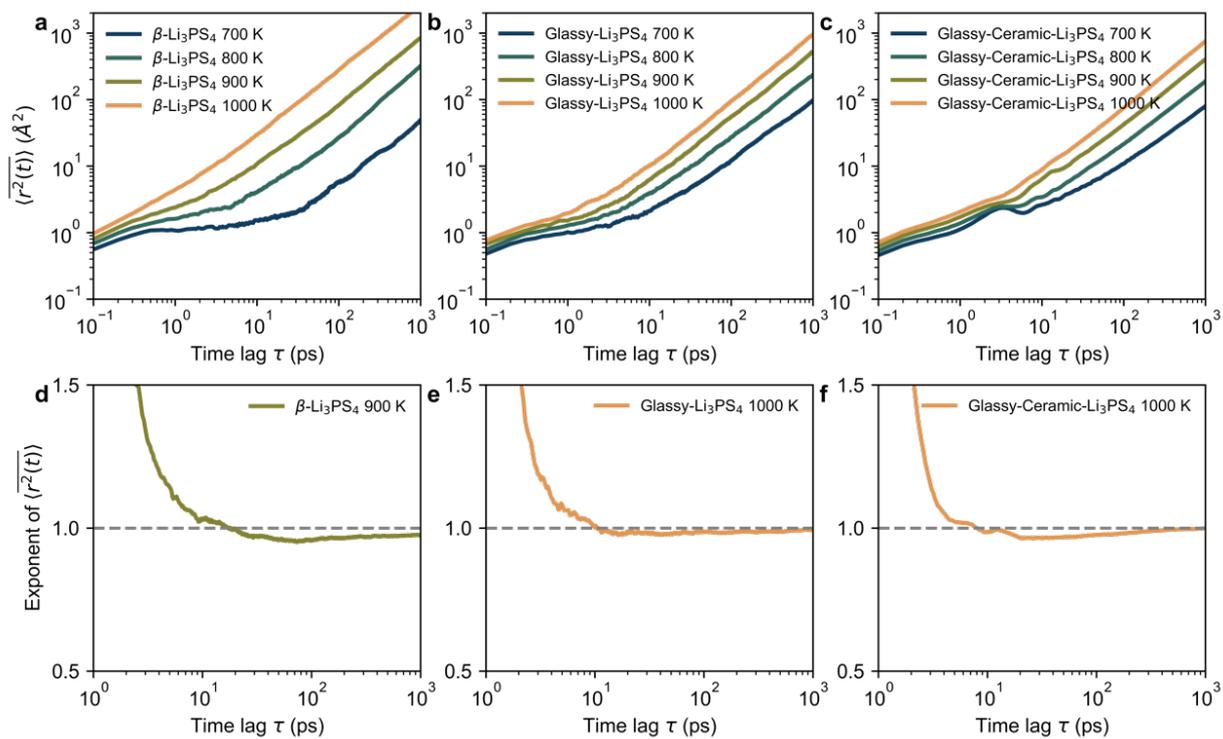

**Fig. S5. a-c** Mean squared displacement of $Li_3PS_4$ electrolytes at different temperature. **d-f** Exponent of mean squared displacement. The horizontal dashed lines represent the Fickian limit $t^1$.

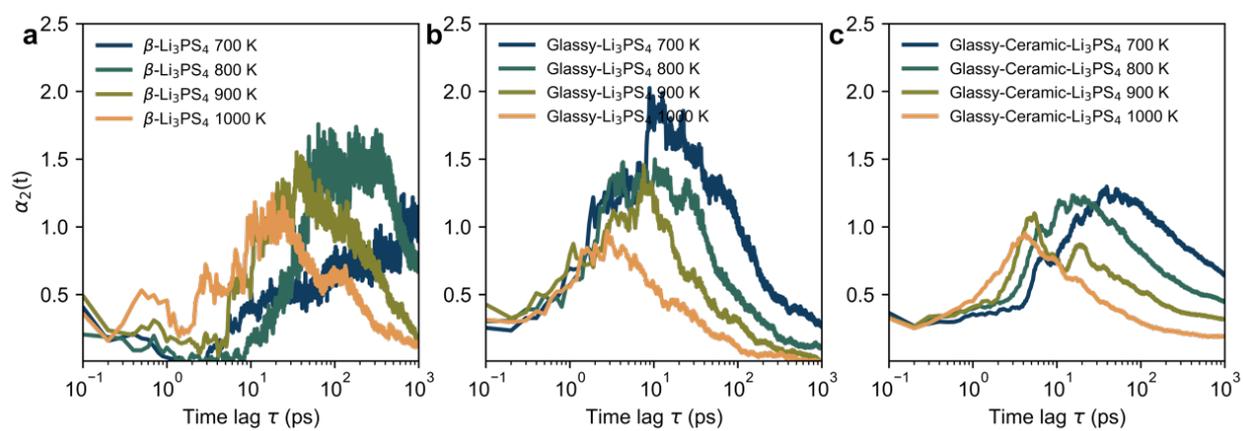

**Fig. S6.** Non-Gaussian parameter of $Li_3PS_4$ electrolytes at different temperatures (from 700 K to 1000 K).

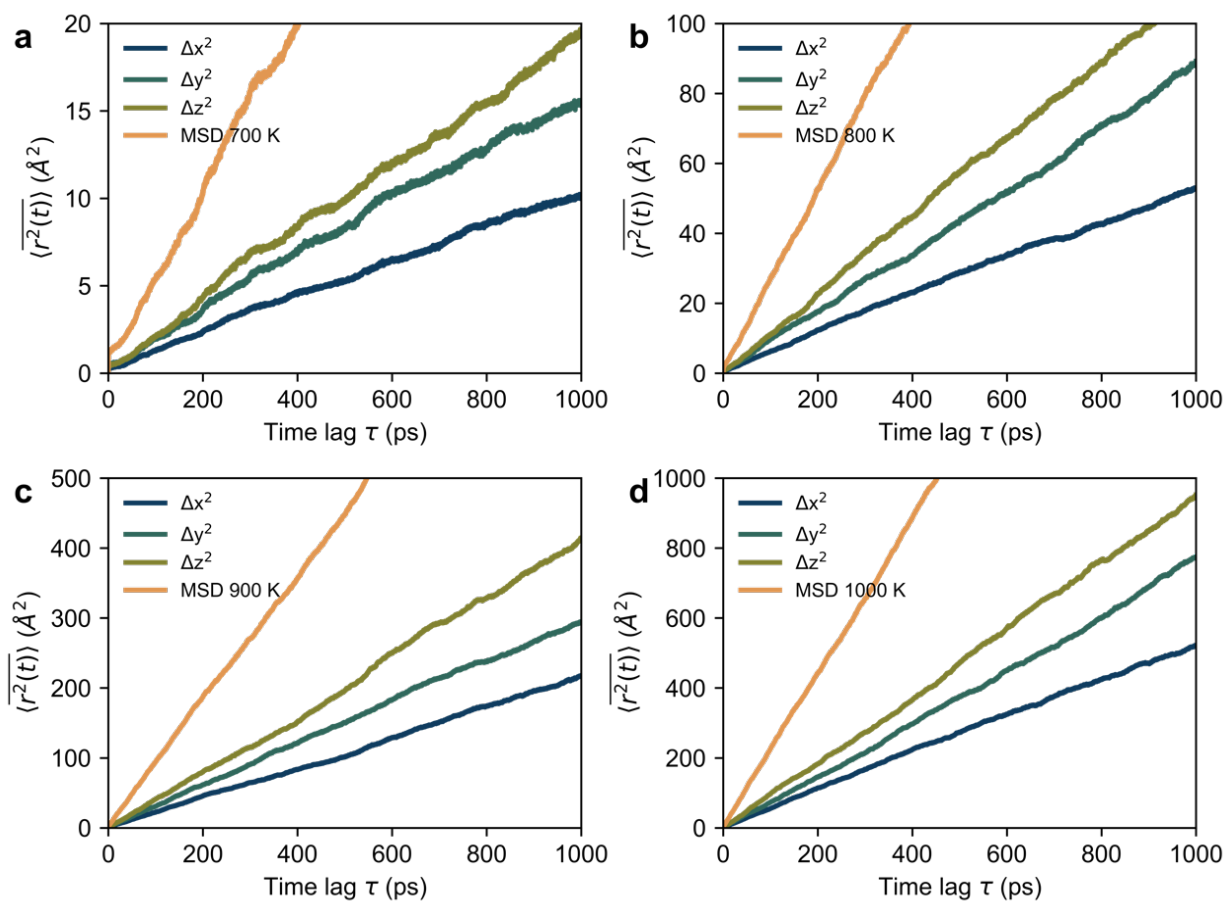

**Fig. S7.** Components of mean squared displacement of β- $Li_3PS_4$ in different directions, where panels **a** to **d** represent mean squared displacement at temperatures ranging from 700 K to 1000 K.

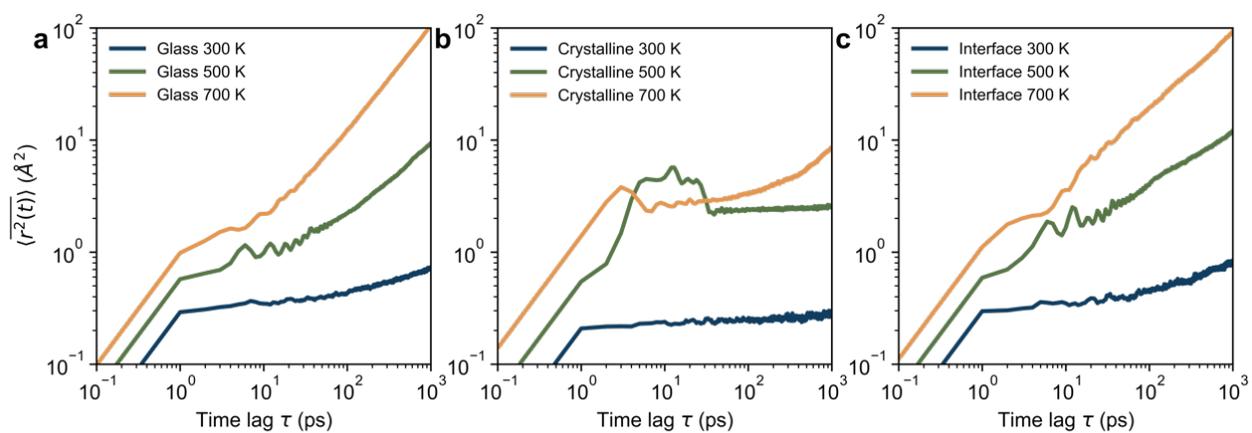

**Fig. S8.** Mean squared displacement of **a** glass, **b** crystalline, and **c** interface phases within the simulated glass-ceramic $Li_3PS_4$ electrolytes at different temperatures.

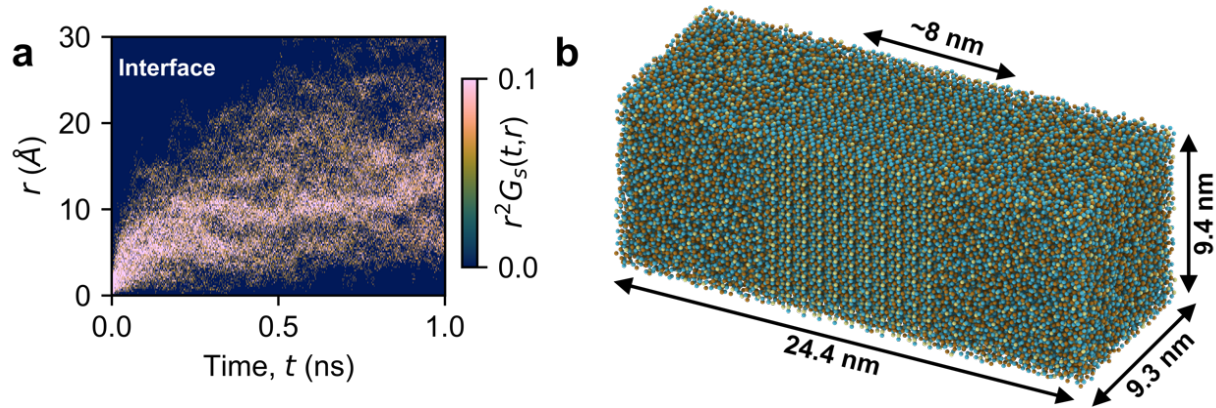

**Fig. S9. a** Self-part van Hove correlation function of interfacial phases within glass-ceramic $Li_3PS_4$ at 900 K. **b** Atomic snapshot of simulated glass-ceramic $Li_3PS_4$, with annotated dimensions representing the sizes of the crystalline and glassy phases.

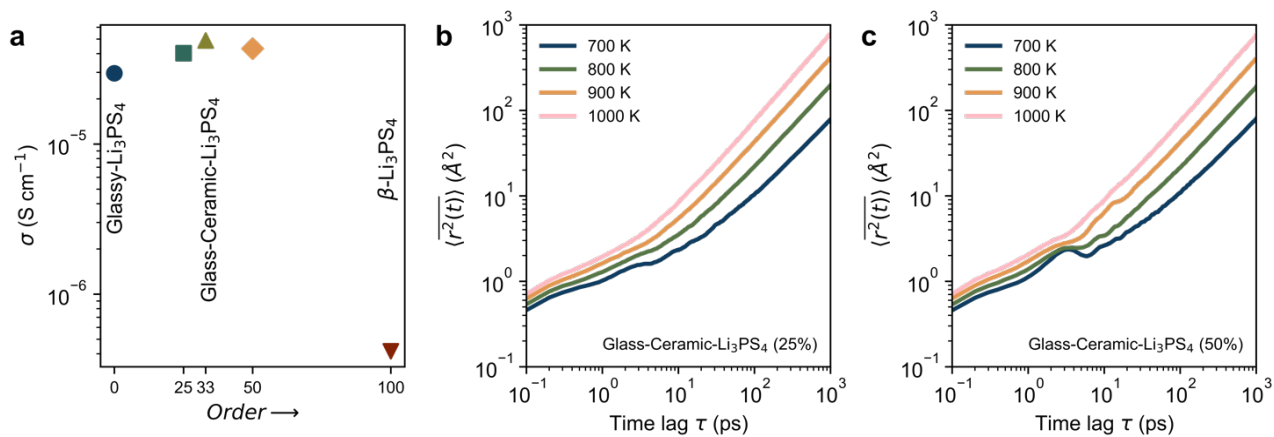

**Fig. S10. a** Room temperature ionic conductivities of glassy, glass-ceramic Li$_3$PS$_4$, and β-Li$_3$PS$_4$ electrolytes. The horizontal axis represents the fraction (%) of crystalline content. **b,c** Mean squared displacement of glass-ceramic Li$_3$PS$_4$ electrolytes with **b** 25% and **c** 50% crystalline content.

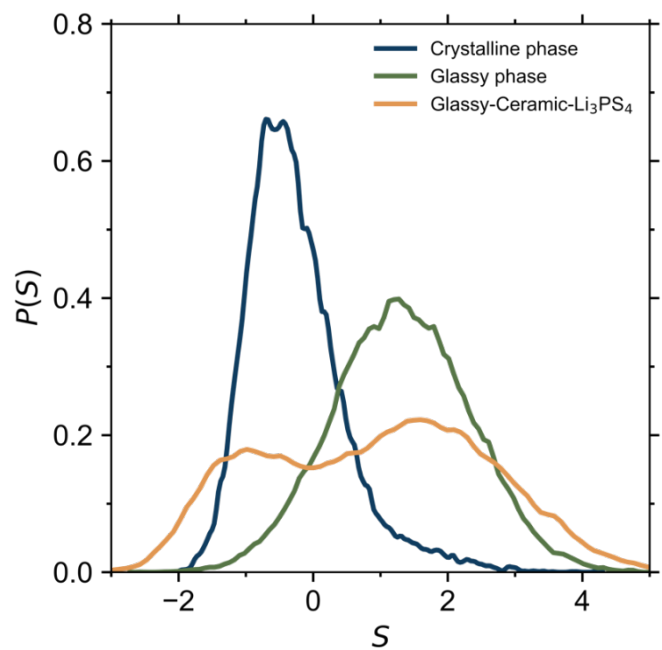

**Fig. S11.** Distribution of lithium softness $S$ for crystalline and glassy phases within the glass-ceramic $Li_3PS_4$ as well as the average of the glass-ceramic $Li_3PS_4$.

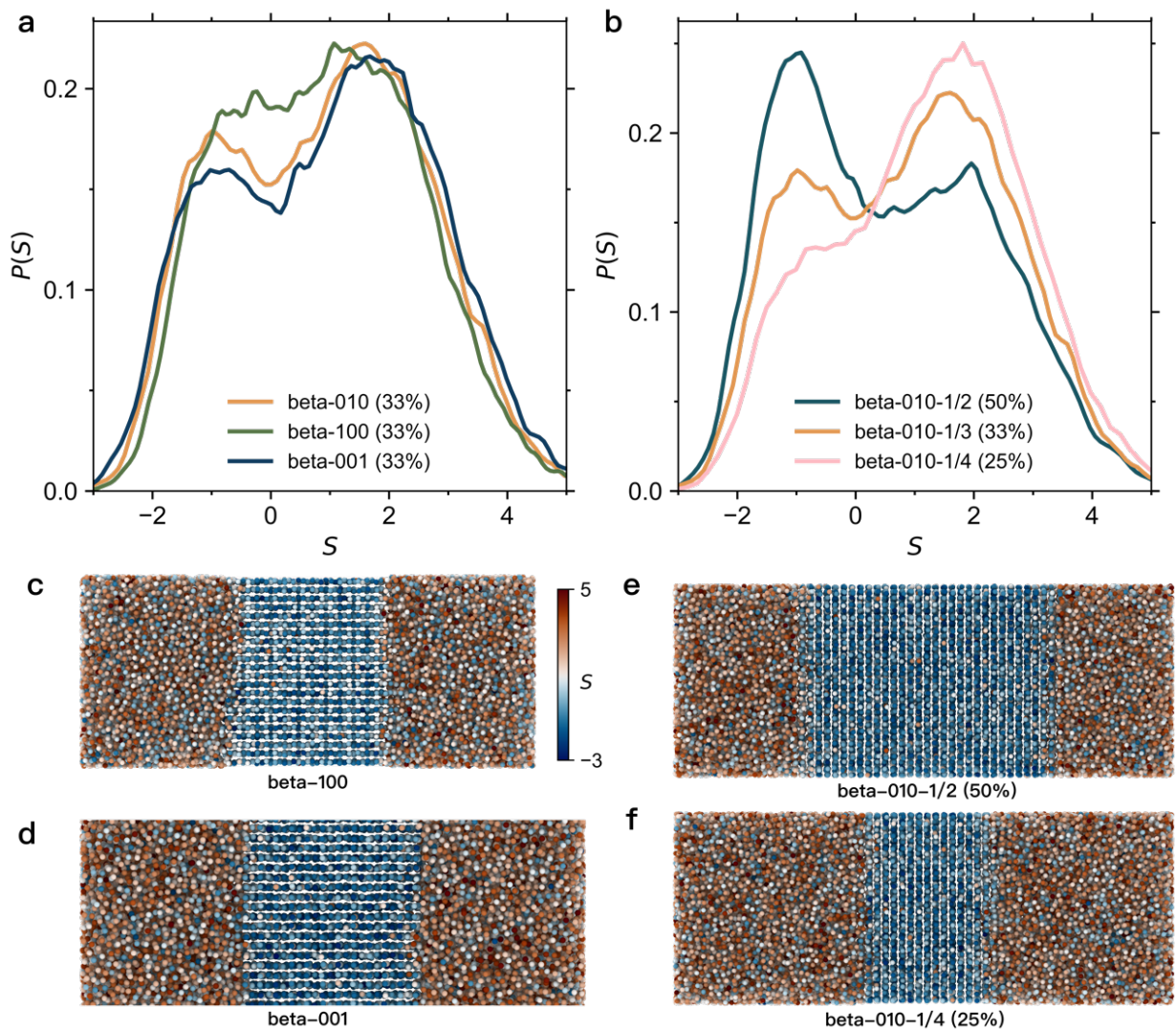

**Fig. S12. a,b** Distribution of particle softness $S$ for the glass-ceramic Li$_3$PS$_4$ with **a** different crystal (β-Li$_3$PS$_4$) orientations and **b** varying crystal content. **c-f** Corresponding atomic snapshots, where only lithium atoms are presented and colored according to their particle softness $S$ value.

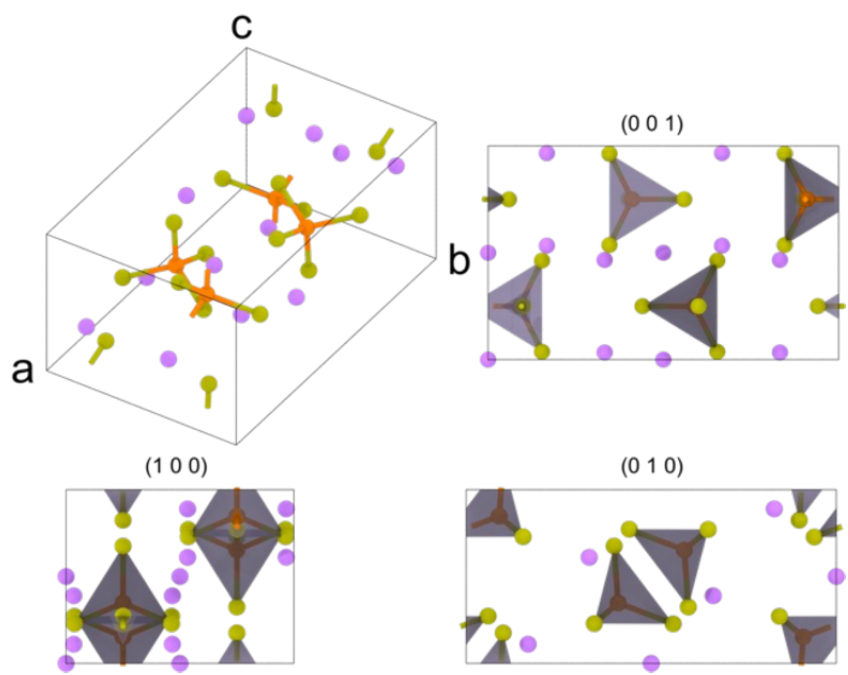

**Fig. S13.** Visual representation of the β-Li₃PS₄ unit cell, highlighting the crystalline planes of (0 0 1), (1 0 0), and (0 1 0). The space group of β-Li₃PS₄ is *Pnma*. Lattice constants: a = 13.03282 Å, b = 8.01860 Å, c = 6.17177 Å, α = 90, β = 90, γ = 90.

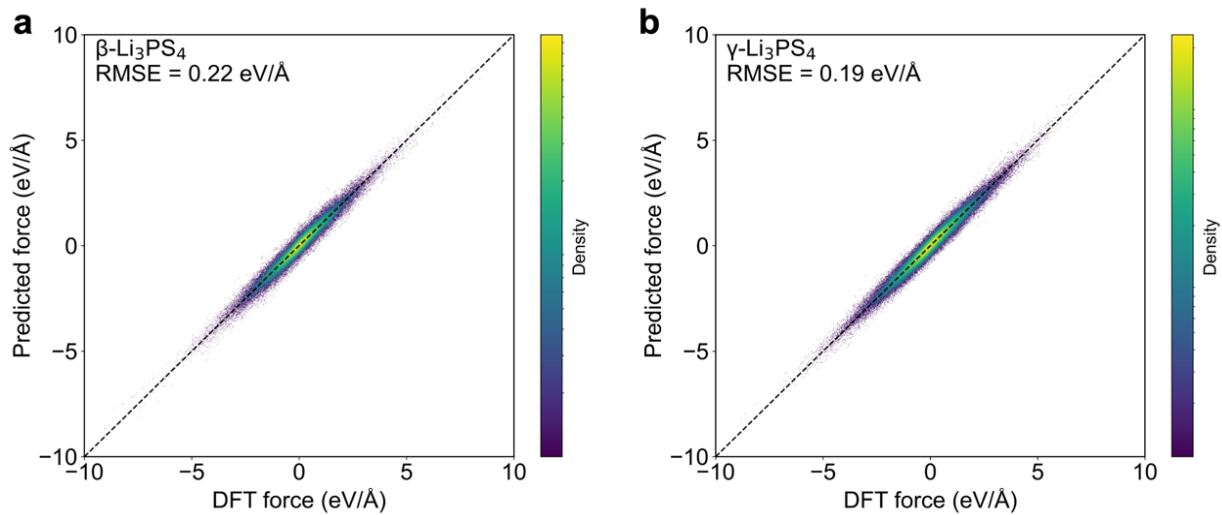

**Fig. S14.** Comparison of MLIP predicted and DFT calculated atomic forces for **a** β-Li$_3$PS$_4$ and **b** γ-Li$_3$PS$_4$ systems during the melting process.

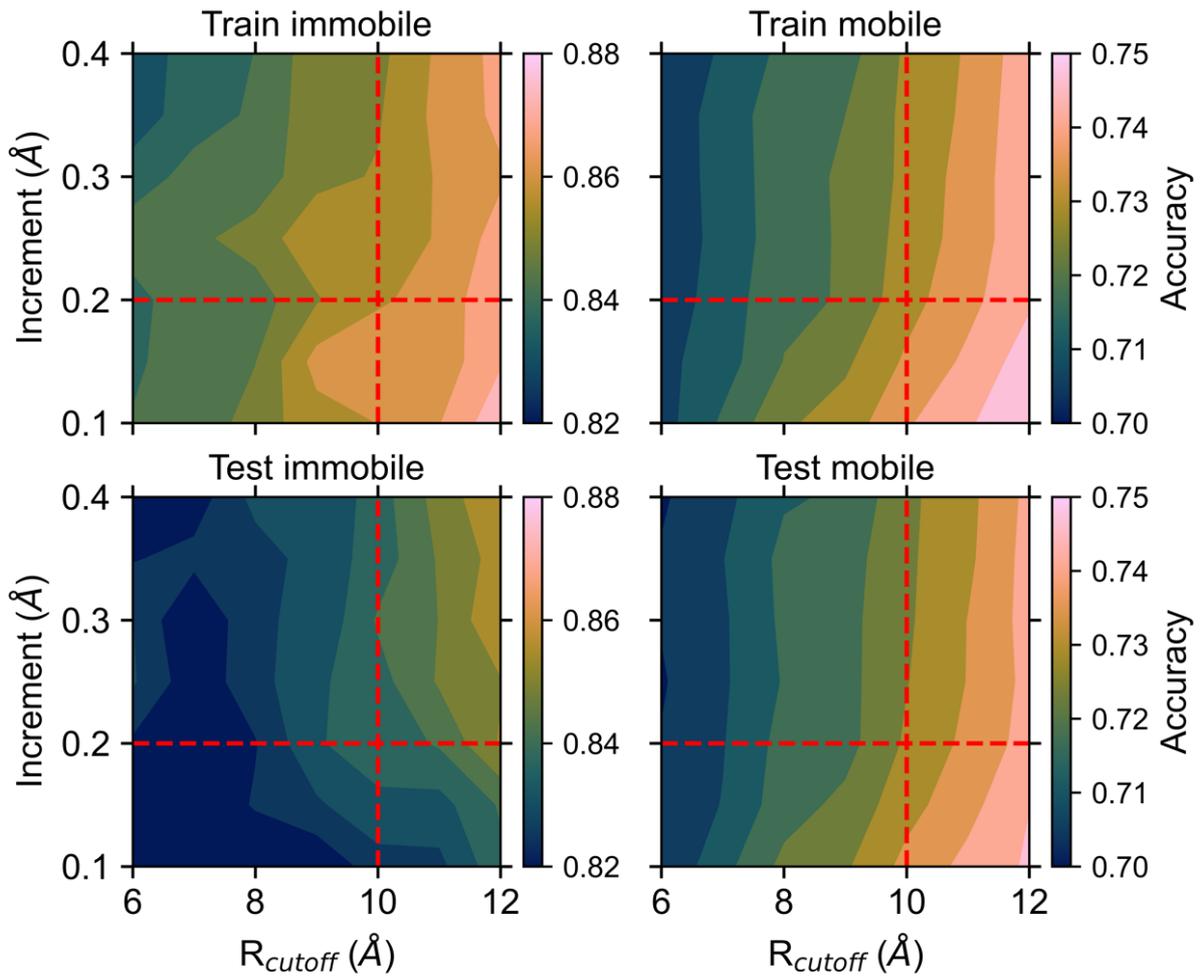

**Fig. S15.** Classification accuracy of the logistic regression model as a function of the *dr* and $R_{cutoff}$ values of the radial order parameters to discriminate the mobile and immobile lithium atoms.

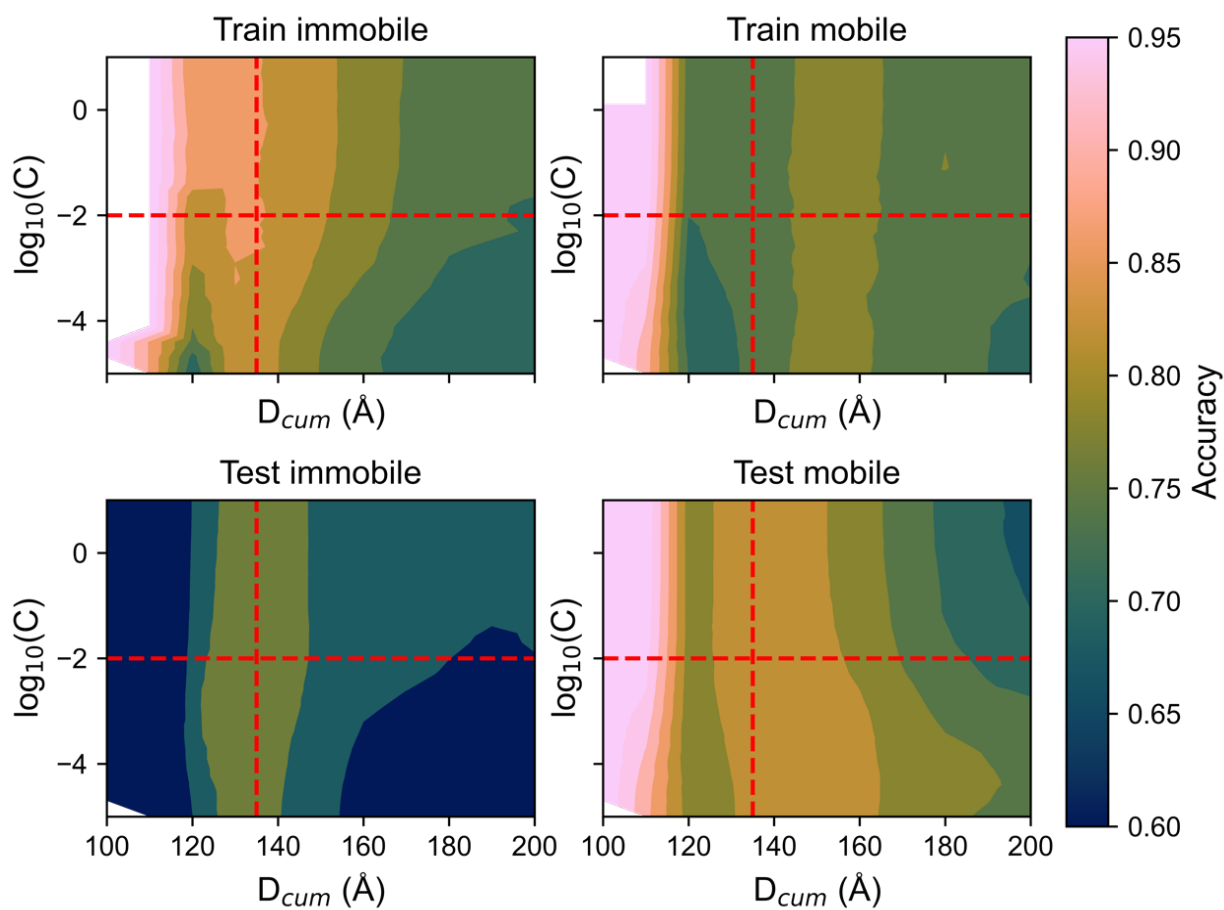

**Fig. S16.** Classification accuracy of the logistic regression model as a function of the regularization parameter $C$ and cumulative non-affine displacement threshold $D_c$ to discriminate the mobile and immobile lithium atoms.

**Supporting Table**

**Table S1.** Details of the initial data set obtained from the *ab initio* MD simulations and used to train the MLIP.

| System | Composition | Atom number | Number of configurations |
|---|---|---|---|
| β-Li$_3$PS$_4$ | Li24P8S32 | 64 | 1000 |
| γ-Li$_3$PS$_4$ | Li48P16S64 | 128 | 1000 |
| Li$_2$P$_2$S$_6$ | Li16P16S48 | 80 | 1000 |
| Hexagonal Li$_2$PS$_3$ | Li32P16S48 | 96 | 1000 |
| Orthorhombic Li$_2$PS$_3$ | Li32P16S48 | 96 | 1000 |
| Li$_2$S | Li64S32 | 96 | 1000 |
| Li$_3$P | Li48P16 | 64 | 1000 |
| Li$_4$P$_2$S$_6$ | Li32P16S48 | 96 | 1000 |
| Li$_7$P$_3$S$_{11}$ | Li28P12S44 | 84 | 1000 |
| Li$_7$PS$_6$ | Li28P4S24 | 56 | 1000 |
| Li$_{48}$P$_{16}$S$_{61}$ | Li48P16S61 | 125 | 1000 |
| P$_2$S$_5$ | P8S20 | 28 | 1000 |
| P$_4$S$_3$ | P32S24 | 56 | 1000 |
| 67Li$_2$S-33P$_2$S$_5$ | Li82P40S138 | 260 | 1000 |
| 70Li$_2$S-30P$_2$S$_5$ | Li82P38S133 | 253 | 1000 |
| 75Li$_2$S-25P$_2$S$_5$ | Li91P35S129 | 255 | 1000 |
| 80Li$_2$S-20P$_2$S$_5$ | Li92P34S128 | 254 | 1000 |
| Li | Li54 | 54 | 1000 |
| P | P48 | 48 | 1000 |
| S | S32 | 32 | 1000 |

**Table S2.** Details of the explored data set obtained from the single energy calculation using CP2K.

| Iteration | System | Composition | Atom number | Conditions | Number of configurations |
|---|---|---|---:|---|---:|
| 1 | 67Li2S-33P2S5 | Li82P40S138 | 260 | Temp(K): | 579 |
|  | 70Li2S-30P2S5 | Li82P38S133 | 253 | [500,800,1000] | 425 |
|  | 75Li2S-25P2S5 | Li91P35S129 | 255 | Press(bar):[0,50] | 159 |
| 2 | 67Li2S-33P2S5 | Li82P40S138 | 260 | Temp(K): | 2344 |
|  | 70Li2S-30P2S5 | Li82P38S133 | 253 | [500,800,1000] | 2309 |
|  | 75Li2S-25P2S5 | Li91P35S129 | 255 | Press(bar):[10,1000] | 802 |
| 3 | 67Li2S-33P2S5 | Li82P40S138 | 260 | Temp(K): | 96 |
|  | 70Li2S-30P2S5 | Li82P38S133 | 253 | [500,800,1000] | 63 |
|  | 75Li2S-25P2S5 | Li91P35S129 | 255 | Press(bar):[5000,10000] | 23 |
| 4 | 67Li2S-33P2S5 | Li82P40S138 | 260 | Temp(K): | 1404 |
|  | 70Li2S-30P2S5 | Li82P38S133 | 253 | [900,1200,1500] | 1459 |
|  | 75Li2S-25P2S5 | Li91P35S129 | 255 | Press(bar):[0,50] | 909 |
| 5 | 67Li2S-33P2S5 | Li82P40S138 | 260 | Temp(K): | 1492 |
|  | 70Li2S-30P2S5 | Li82P38S133 | 253 | [900,1200,1500] | 1495 |
|  | 75Li2S-25P2S5 | Li91P35S129 | 255 | Press(bar):[10,1000] | 866 |
| 6 | 67Li2S-33P2S5 | Li82P40S138 | 260 | Temp(K): | 1492 |
|  | 70Li2S-30P2S5 | Li82P38S133 | 253 | [900,1200,1500] | 1489 |
|  | 75Li2S-25P2S5 | Li91P35S129 | 255 | Press(bar):[5000,10000] | 756 |
| 7 | 67Li2S-33P2S5 | Li82P40S138 | 260 | Temp(K): | 1498 |
|  | 70Li2S-30P2S5 | Li82P38S133 | 253 | [900,1200,1500] | 1490 |
|  | 75Li2S-25P2S5 | Li91P35S129 | 255 | Press(bar):[0,50] | 1491 |
| 8 | 67Li2S-33P2S5 | Li82P40S138 | 260 | Temp(K): | 1498 |
|  | 70Li2S-30P2S5 | Li82P38S133 | 253 | [1400,1600,1800] | 1495 |
|  | 75Li2S-25P2S5 | Li91P35S129 | 255 | Press(bar):[10,1000] | 1490 |
| 9 | 67Li2S-33P2S5 | Li82P40S138 | 260 | Temp(K): | 1497 |
|  | 70Li2S-30P2S5 | Li82P38S133 | 253 | [1400,1600,1800] | 1498 |
|  | 75Li2S-25P2S5 | Li91P35S129 | 255 | Press(bar):[5000,10000] | 1491 |
| 10 | 67Li2S-33P2S5 | Li82P40S138 | 260 | Temp(K): | 1490 |
|  | 70Li2S-30P2S5 | Li82P38S133 | 253 | [400,800,1200] | 1497 |
|  | 75Li2S-25P2S5 | Li91P35S129 | 255 | Press(bar):[100] | 1491 |
| 11 | 67Li2S-33P2S5 | Li82P40S138 | 260 | Temp(K): | 1488 |
|  | 70Li2S-30P2S5 | Li82P38S133 | 253 | [400,800,1200] | 1489 |
|  | 75Li2S-25P2S5 | Li91P35S129 | 255 | Press(bar):[2000] | 1250 |